\documentclass[11pt]{article}
\pdfoutput=1
\usepackage{jheppub}

\usepackage{amsfonts}
\usepackage{amscd}
\usepackage{amssymb}
\usepackage{amsmath,bbm}
\usepackage{epsfig}
\usepackage{latexsym}
\usepackage{mathtools}
\usepackage{arydshln}
\usepackage{hyperref}
\usepackage{color}
\usepackage{slashed}
\usepackage{euscript, mathrsfs}
\usepackage[vcentermath]{youngtab}
\usepackage{marginnote}
\usepackage{bm}

\usepackage{comment}

\def\be{\begin{equation}}
\def\ee{\end{equation}}
\def\bea{\begin{eqnarray}}
\def\eea{\end{eqnarray}}
\def \beaa {\begin{equation}\begin{aligned}}
\def \eeaa {\end{aligned}\end{equation}}

\newcommand{\nn}{\nonumber}

\newcommand\diff{\mathrm{d}}

\newcommand{\ii}{\mathrm{i}}

\newcommand\g{\gamma}
\newcommand\e{\epsilon}
\newcommand\G{\Gamma}

\newlength{\sswidth}

\def\fp{\mathfrak{p}}
\def\fq{\mathfrak{q}}
\def\ft{\mathfrak{t}}

\def\a{\alpha}
\def\b{\beta}
\def\g{\gamma}

\def\e{\epsilon}

\def\CV{{\cal V}}
\def\CF{{\cal F}}

\def\IC{\mathbb{C}}

\def\IZ{\mathbb{Z}}

\def\half{\frac{1}{2}}
\def\p{\partial}

\newcommand{\ket}[1]{\left|{#1}\right\rangle}

\def\vec#1{\bm{#1}}

\def \be  {\begin{equation}}
\def \ee  {\end{equation}}
\def \ba  {\begin{eqnarray}}
\def \ea  {\end{eqnarray}}
\def \bb  {\begin{thebibliography}}
\def \eb  {\end{thebibliography}}
\def \lab #1 {\label{#1}}

\newcommand\cA{\mathcal{A}}

\newcommand\cH{\mathcal{H}}
\newcommand\cI{\mathcal{I}}
\newcommand\CI{\mathcal{I}}

\newcommand\cK{\mathcal{K}}

\newcommand\cN{\mathcal{N}}
\newcommand\CN{\mathcal{N}}
\newcommand\cO{\mathcal{O}}
\newcommand\cQ{\mathcal{Q}}

\newcommand\cS{\mathcal{S}}
\newcommand\cT{\mathcal{T}}

\newcommand\cW{\mathcal{W}}

\newcommand\tr{\mathrm{Tr}}

\newcommand\ie{\textit{i.e.}}

\title{Four-dimensional Lens Space Index from Two-dimensional Chiral Algebra}
\author[a]{Martin Fluder}
\author[b,c]{and Jaewon Song}
\affiliation[a]{
Walter Burke Institute for Theoretical Physics,
California Institute of Technology\\
Pasadena, CA 91125, USA
}
\affiliation[b]{School of Physics, Korea Institute for Advanced Study, Seoul 02455, Korea}
\affiliation[c]{Department of Physics, University of California, San Diego, 
La Jolla, CA 92093, USA}

\emailAdd{fluder@caltech.edu}
\emailAdd{jsong@kias.re.kr}

\preprint{CALT-TH-2017-037, KIAS-P17051}

\abstract{
We study the supersymmetric partition function on $S^1 \times L(r, 1)$, or the lens space index of four-dimensional $\cN=2$ superconformal field theories and their connection to two-dimensional chiral algebras. We primarily focus on free theories as well as Argyres-Douglas theories of type $(A_1, A_k)$ and $(A_1, D_k)$. We observe that in specific limits, the lens space index is reproduced in terms of the (refined) character of an appropriately twisted module of the associated two-dimensional chiral algebra or a generalized vertex operator algebra. The particular twisted module is determined by the choice of discrete holonomies for the flavor symmetry in four-dimensions.
}

\begin{document}
\maketitle

\bibliographystyle{JHEP}


\section{Introduction}\label{SecIntroduction}

Recently, it was discovered that four-dimensional $\cN=2$ superconformal field theories (SCFT) have protected subsectors described by two-dimensional chiral algebras or vertex operator algebras (VOAs)~\cite{Beem:2013sza}. This relation defines a map (or functor) from the set of four-dimensional $\CN=2$ superconformal field theories to the set of two-dimensional chiral algebras, or more precisely the Vertex Operator Algebras (VOA), 
\begin{align}
 \chi: \{ \textrm{4d }\cN=2\textrm{ SCFT} \}/\left( \textrm{marginal deformations} \right) \to \{ \textrm{2d chiral algebra (VOA)} \} \, . 
\end{align}
Under this map, various observables of the four-dimensional theory, which are invariant under exactly marginal deformations, are mapped to quantities in the two-dimensional chiral algebra~\cite{Song:2016yfd, Cordova:2017mhb, Beem:2017ooy}. Aspects of chiral algebras arising from four-dimensional $\cN=2$ superconformal field theories have been discussed in~\cite{Beem:2014kka,Beem:2014rza,Lemos:2014lua, Lemos:2015orc,Liendo:2015ofa,Arakawa:2016hkg,Xie:2016evu,Buican:2016arp,Neitzke:2017cxz,Song:2017oew, Creutzig:2017qyf, Fredrickson:2017yka}. 
In the following we give a list of observables, which are captured by the protected two-dimensional chiral algebra: 
\vspace{.2cm}
\begin{align}
{\renewcommand{\arraystretch}{1.4}
	\begin{array}{|c|c|}
	\hline
	 \qquad\qquad 4d~\cN=2~ \textrm{SCFT}\qquad\qquad& \qquad\qquad2d~\textrm{chiral algebra}\qquad\qquad \\
	 \hline \hline
	 \textrm{ superconformal symmetry } & \textrm{Virasoro symmetry} \\
	 SU(2)_R \textrm{ current}  & \textrm{stress tensor} \\
	 \textrm{conformal anomaly } c_{4d} & \textrm{central charge } c_{2d} = -12 c_{4d} \\
	 \hline
	 \textrm{global symmetry } \mathfrak{g} & \textrm{affine symmetry } \hat{\mathfrak{g}}_{k_{2d}} \\
	 \textrm{moment map operator} & \textrm{conserved current} \\
	 \textrm{flavor central charge } k_{4d} & \textrm{level } k_{2d} = -\half k_{4d} \\
	 \hline
	 \textrm{Schur index} & \textrm{vacuum character} \\
	 \textrm{Higgs branch} & \textrm{associated variety} \\
	 \textrm{Surface defect} & \textrm{non-vacuum module} \\
	 \hline
	 \textrm{Macdonald index} & \textrm{`refined' vacuum character} \\
	 \textrm{lens space index} & \textrm{``twisted character"} \\
	 \hline
	\end{array}}\nn
\end{align}
The purpose of the current paper is to add the last line to this already rather extensive dictionary. 

The superconformal index of an $\CN=2$ SCFT admits various simplifying limits \cite{Gadde:2011ik,Gadde:2011uv}. The ``Schur limit" of the superconformal index is of particular interest in the four-dimensional $\cN=2$ SCFT/two-dimensional chiral algebra correspondence, since it precisely matches to the vacuum character of the chiral algebra. The ``Macdonald limit" of the index counts the same set of $\frac{1}{4}$-BPS operators as the Schur index. However, it has one more variable and therefore carries more \emph{refined} information than the Schur index. A way to compute the Macdonald limit of the index from the chiral algebra has been proposed in~\cite{Song:2016yfd}. It is interesting because the supercharge used to define the cohomological construction of the chiral algebra is \emph{not} compatible with the extra grading needed to define the Macdonald index. Nevertheless it was possible (at least for simple cases) to define a graded vector space associated to the chiral algebra capturing the Macdonald index, thanks to a filtration structure in the vacuum module for the chiral algebra (or VOA)~\cite{feigin2009pbw,Li2005}. 

The goal of the current paper is to identify a two-dimensional quantity that captures the lens space index~\cite{Benini:2011nc}, or -- equivalently -- the supersymmetric partition function on $S^1 \times S^3/\IZ_r$. We give a definition and detailed review of the lens space index in Appendix~\ref{app:LensIdx}. One first important feature of the lens index is that one can turn on discrete holonomy for background gauge fields of the flavor symmetry, which will partially break the flavor symmetry group. Thus the lens index is parametrized by such a choice of holonomy, which we shall collectively denote as $\vec{m}$. Upon gauging the flavor symmetry group, we are supposed to sum over all possible choices of the discrete holonomy. Another important feature of the lens index is that there are zero-point energy contributions to each matter multiplet. As we shall see, we will have to first strip off the overall zero-point energy contribution, in order to properly take the Macdonald limit of the lens index. 

The orbifold-prescription, which is being imposed at the level of the lens space $L(r, 1)=S^3/\IZ_r$, can be translated into the two-dimensional chiral algebra by taking a quotient on the chiral algebra plane $\IC/\IZ_r$. Schematically, this will lead to the following condition on the fields in the chiral algebra 
\bea
 \phi \left(e^{\frac{2\pi \ii }{r}} z\right) \ = \ g \, \phi\left(z\right) \,,
\eea
where $g$ is an element of the symmetry group (or automorphism of the vertex operator algebra) of the theory, such that $g^r=1$. This can be thought of as a twisted sector upon mapping $\IC/\IZ_r \to \IC$, via $z \mapsto z^r$. This map acts on the Virasoro generators as $L_n \mapsto r L_n$, which in turn has the effect of rescaling $q \to q^r$. We will be required to account for that rescaling when we compute the twisted character of the chiral algebra and match it with the lens space index. 

Similar to the Macdonald limit of the round superconformal index, we observe that the supercharge
\begin{align}
\mathbbmtt{Q} \ = \ \cQ^1_{-}+\tilde \cS^{\dot{-}2}
\end{align}
used to define the chiral algebra in~\cite{Beem:2013sza} is \emph{not} compatible with the $\IZ_r$ action on the chiral algebra plane. Nevertheless we shall explicitly observe that the lens space index can be associated with a character of the chiral algebra defined via the supercharge $\mathbbmtt{Q}$.\footnote{A similar phenomenon was demonstrated in \cite{Cordova:2016uwk}. There, the line operator Schur index was found to be written in terms of the characters of the associated chiral algebra even though the BPS line operator is does not preserve the supercharge $\mathbbmtt{Q}$.}  

Indeed, we find that the (refined) character of a twisted module $\CV^{(\vec{\lambda})}$ precisely reproduces the lens index in the Macdonald limit:
\begin{align}
\chi^{\textrm{ref}}_{\CV^{(\vec{\lambda})}} (\vec{a}; q, T) \ = \ 
     \begin{dcases} 
     \CI^{\textrm{Mac}}_{S^3/\IZ_r} (\vec{a}, \vec{m} \neq 0; q \to q^{1/r}, t \to T) \ ,  & \quad (\vec{\lambda} \neq 0) \,, \\
     \CI^{\textrm{Mac}}_{S^3/\IZ_r} (\vec{a}, \vec{m} = 0; q \to q^{1/r}, t \to qT) \ ,  & \quad  (\vec{\lambda} = 0) \,,
     \end{dcases}
\end{align} 
where the twist parameter $\vec{\lambda}$ in the chiral algebra depends on the choice of holonomy $\vec{m}$. One immediate point to notice is that one should take two \emph{distinct} limits of the four-dimensional lens index to match with the two-dimensional chiral algebra. Similarly, we observe that the left-hand side of the correspondence is given in terms of the refined character as defined in \cite{Song:2016yfd}. Typically, when we take $T \to 1$, the refined character becomes the usual character. However in some cases, the $T\to1$ limit (or $t=q^r$ limit of the lens index) diverges, and in these cases, the correspondence works \emph{only} if we keep the extra fugacity of the refined character. 

The twisted module $\CV^{(\vec{\lambda})}$ we define here is quite different from what is commonly known as a twisted modules. Firstly, depending on the twist parameter $\lambda$, the highest-weight state carries non-zero eigenvalues of the flavor current algebra. Secondly, the highest-weight state can be an eigenstate of the zero modes of the current generator. These features are unusual, but we verify that these conditions precisely give us the lens index of the four-dimensional superconformal field theory in question. 

In this paper, we will mainly focus on the ``simplest" four-dimensional $\cN=2$ superconformal field theories, namely free theories and Argyres-Douglas (AD) theories~\cite{Argyres:1995jj,Argyres:1995xn, Eguchi:1996vu, Xie:2012hs, Wang:2015mra}. Similarly, the corresponding chiral algebras~\cite{Cordova:2015nma, Liendo:2015ofa, Xie:2016evu, Creutzig:2017qyf} are rather straightforward to work with. The (round sphere) superconformal indices for the (generalized) Argyres-Douglas theories have recently been computed using various independent methods.  
We will explicitly employ the recently found four-dimensional $\cN=1$ Lagrangians, which were shown to flow to Argyres-Douglas theory in~\cite{Maruyoshi:2016tqk, Maruyoshi:2016aim, Agarwal:2016pjo,Agarwal:2017roi,Benvenuti:2017bpg}, to compute the lens index of such $\cN=2$ theories. In particular, using such $\cN=1$ Lagrangians to compute the round Schur or Macdonald indices were explicitly shown to agree with other methods, such as employing topological field theory arguments~\cite{Buican:2015hsa,Buican:2015ina,Buican:2015tda,Song:2015wta,Buican:2017uka,Song:2017oew}, or BPS monodromies~\cite{Cordova:2015nma,Cecotti:2015lab, Cordova:2016uwk,Cordova:2017mhb,Cordova:2017ohl}.

This paper is organized as follows. In Section~\ref{SecFreeCA}, we explicitly work out the correspondence between the lens index and two-dimensional chiral algebra twisted characters for free four-dimensional $\cN=2$ superconformal field theories. In Section~\ref{sec:ADindex}, we define and compute the lens index for four-dimensional Argyres-Douglas theories using the $\cN=1$ Lagrangians introduced in~\cite{Maruyoshi:2016tqk,Maruyoshi:2016aim,Agarwal:2016pjo}. In Section~\ref{SecAD}, we work out the twisted characters for some examples of affine Kac-Moody algebras, and their precise relation to lens indices arising from the corresponding four-dimensional Argyres-Douglas theories. Finally in Section~\ref{SecDiscussion}, we conclude by discussing our results and mentioning a few interesting future directions. In Appendix~\ref{app:LensIdx}, we review the definition of the lens index in detail. 


\section{Twisted chiral algebras for free theories}\label{SecFreeCA}

In this Section we identify the chiral algebra corresponding to the lens index of free four-dimensional $\cN=2$ superconformal field theories. We start by treating the $\beta-\gamma$ system, which corresponds to a free four-dimensional $\cN=2$ hypermultiplet, and then discuss the $(b,c)$ ghost system, corresponding to a free four-dimensional $\cN=2$ vector multiplet.

\subsection{The twisted $\beta-\gamma$ system and the free hypermultiplet}

The chiral algebra for the free $\cN=2$ hypermultiplet is given by the system of symplectic bosons or equivalently a $\beta-\gamma$ system with $(h_\beta, h_\gamma)=(\half, \half)$. Aspects of this theory are discussed in detail, for example in~\cite{Eholzer:1997se,Lesage:2002ch}. Let us first write down some generalities about the chiral algebra corresponding to the free hypermultiplet:

\begin{itemize}
\item The chiral algebra is generated by two fields: $\beta(z)$ and $\gamma(z)$. 
\item The OPE between the fields $\beta$ and $\gamma$ are given as: 
\bea
\beta\left( z \right) \gamma \left( 0 \right) \ = \ - \gamma \left( z \right)\beta\left( 0 \right) \ \sim \ \frac{1}{z} \,.
\eea
\item The stress-energy tensor reads
 \begin{align}
  T  \ = \ \frac{1}{2} \left( : \beta \partial \gamma : - : \gamma \partial \beta : \right) \, ,
 \end{align} 
 where $::$ denotes the usual normal ordering. 
 It follows that the theory has $c=-1$ and both $\beta$ and $\gamma$ have weight $\frac{1}{2}$. 
 \item There is are $\mathfrak{su}(2)$ currents generated by $\beta^2, \beta \gamma$, and $\gamma^2$. They generate the affine Kac-Moody algebra $\widehat{\mathfrak{su}}(2)_{-\half}$. 
\item The vacuum character of this chiral algebra is simply given by 
 \begin{align}
  \chi_{\CV} (a; q) \ =  \ \tr_{\CV} q^{L_0} a^{J_3} \ = \ \frac{1}{(q^\half a; q)(q^\half a^{-1}; q)} \, , 
 \end{align}
 where $(x; q) = \prod_{i=0}^{\infty} (1-xq^i)$ is the $q$-Pochammer symbol. This is identical to the Schur index of a free $\cN=2$ hypermultiplet in four dimensions. 
\end{itemize}

Let us now consider a twisted module for the $\beta-\gamma$ system. We introduce the following twisted boundary conditions for the fields
\bea
 \beta(e^{2\pi i} z)  \ = \  e^{-2\pi i \a} \beta(z) \, , \qquad \gamma(e^{2 \pi i}z)  \ = \  e^{2\pi i \a} \gamma(z) \, , 
\eea
This will not change the boundary condition for the stress-energy tensor nor will it affect the OPEs of the $\beta-\gamma$ system, however it will break the $\mathfrak{su}(2)$ symmetry down to $\mathfrak{u}(1)$. 

Given the above ``twisting action", we may write down the mode expansions of the fields as follows (the subscript for each modes is chosen in such a way that it is identical to the $L_0$ weight)
\bea
	\beta(z) =  \sum_{n\in \IZ} \frac{\beta_{n+\a +\half }}{z^{n+\a + 1}} = \sum_{n \in \IZ} \frac{\beta_{n+\lambda}}{z^{n+\half + \lambda}} \,, \qquad
	\gamma(z) = \sum_{n\in \IZ} \frac{\gamma_{n-\a + \half}}{z^{n-\a + 1}} = \sum_{n \in \IZ} \frac{\gamma_{n-\lambda}}{z^{n+\half-\lambda}} \,,
\eea
where we define $\lambda \equiv \a + \half$. The modes satisfies the following algebra
\begin{align}
 [\beta_{n+\a+\half}, \gamma_{m-\a + \half}] \ = \ \delta_{n+m+1, 0} \, , 
\end{align}
or equivalently
\begin{align}
 [\beta_{n+\lambda}, \gamma_{m-\lambda}] \ = \ \delta_{n+m, 0} \, . 
\end{align}
We define the vacuum state as 
\begin{align}
\beta_{a} \ket {0} \ = \ \gamma_{a} \ket{0} \ = \ 0 \,, \qquad \text{for } a \ge 0 \, . 
\end{align}
The character for the twisted vacuum module can be computed by counting the descendant states. 
When $ \lambda \neq 0$ these are
\begin{align}
\begin{split}
 & \beta_{-1+\lambda}\,, ~\beta_{-2+\lambda}\,, ~\beta_{-3+\lambda}\,,~ \cdots \,, \\
 & \gamma_{-\lambda}\,, ~\gamma_{-1-\lambda} \,, ~\gamma_{-2+\lambda}\,,~ \cdots  \,.
\end{split}
\end{align}
Therefore we get the following twisted vacuum character for $\lambda \neq 0$
\begin{align}
 \chi_{\CV^{(\lambda)}} (a; q) \ = \ \frac{1}{ (q^{1-\lambda} a ; q) (q^{\lambda} a^{-1}; q)} \, , 
\end{align}
where the first and second factor comes from the $\beta$-modes and $\gamma$-modes respectively. The fugacity $a$ is for the $U(1)$ global symmetry. 
Without twisting, we set $\a =0$ or $\lambda = \half$. So we see that the untwisted vacuum character is given by
\begin{align}
 \chi_{\CV^{(0)}} (a; q) \ = \ \frac{1}{ (q^{\half} a ; q) (q^{\half} a^{-1}; q)} \, , 
\end{align}
The case with $\lambda = 0$ for the twisted module is special, and the character is given by $(q a; q)^{-1} (q/a; q)^{-1}$. But this case is not connected to the lens index as we shall see later. Therefore, we use the symbol $\CV^{(0)}$ to denote the case of untwisted module ($\a=0$), instead of the case with $\lambda=0$. 

\subsubsection{Refined character} \label{sec:refinedchar}
 One can also consider the refined character~\cite{Song:2016yfd} for the twisted $\beta-\gamma$ system. 
 Let us briefly review the notion of the refined character. The chiral algebra admits an filtration $\CV_0 \subset \CV_{1/2} \subset \CV_{1} \subset \ldots$, given by
 \begin{align} \label{eq:filter}
  \CV_k \ = \ \textrm{span} \left\{ X^{(i_1)}_{-n_1} \cdots X^{(i_m)}_{-n_m} \ket{\Omega} : n_1 \ge \cdots \ge n_m, \sum_{j=1}^m {w(X^{(i_j)})} \le k \right\} \Big/ \left\{\textrm{null states} \right\} \, , 
 \end{align}
where $\Omega$ is the highest-weight state and $X^{(i)}_{-n}$ are the generators of the chiral algebra with the subscript denoting modes in a Laurent expansion. Here $w(X)$ denotes the ``weight" of the generator $X$, that we assign.~\footnote{As of now, it is unknown whether this weight $w(X)$ can be determined purely from the associated chiral algebra without referring to the four-dimensional $\CN=2$ superconformal field theory data.} From this filtration, we can construct an associated graded vector space, given by
\begin{align} 
 V_{gr} \ = \ \bigoplus_{i=0}^{\infty} V_i \ = \ \CV_0 \oplus \bigoplus_{i > 0}^{\infty} \left(\CV_{i}\big/\CV_{i-\half} \right) \, , 
\end{align}
with $V_i = \CV_{i}/\CV_{i-1/2}$, for $i>0$ and $V_0 = \CV_0$. The refined (super-) character is now defined as
\begin{align} 
 \chi^{\textrm{ref}}_{\CV}(q, T) \ = \ \sum_{i\ge0} \tr_{V_i} (-1)^F q^{L_0} T^i \, , 
\end{align}
where $L_n$ are the Virasoro generators. When the underlying four-dimensional theory has a global flavor symmetry, the chiral algebra contains global symmetry generators that commute with the Virasoro algebra. In this case, we write
\begin{align}\label{eq:refChar}
 \chi^{\textrm{ref}}_{\CV}(\vec{z}; q, T) \ = \ \sum_{i\ge0} \tr_{V_i} (-1)^F q^{L_0} \vec{z}^{\vec{F}} T^i \, , 
\end{align}
where $\vec{F}$ denotes the Cartans for the global symmetry. We also use the short-hand notation $\vec{z}^{\vec{F}} \equiv \prod_i z_i^{F_i}$. To summarize, the extra fugacity $T$ counts the number of generators that are acting on a vacuum $\ket{\Omega}$ to create a particular state carrying the quantum numbers $L_0$, and ${\bf F}$. On top of this, one should take care of null states and the relations among the generators. 

The chiral algebra of the $\beta-\gamma$ system is freely generated and there is no null state in the vacuum module $\CV$. By assigning $w(\beta)=w(\gamma) = \half$, we obtain
\begin{align}
 \chi^{\textrm{ref}}_{\CV^{(\lambda)}} (a; q, T) & \ = \  \frac{1}{(T^{\half} q^{1-\lambda}a ; q) (T^{\half} q^{\lambda} a^{-1}; q)} & (\textrm{twisted}) \, , \label{eq:bgTwisted}\\
 \chi^{\textrm{ref}}_{\CV^{(0)}} (a; q, T) & \ = \ \ \frac{1}{(T^{\half} q^{\half} a ; q) (T^{\half} q^{\half} a^{-1}; q)} & (\textrm{untwisted}) \, . \label{eq:bgUntwisted}
\end{align}

\paragraph{Lens index for a free hypermultiplet from the twisted character.}

The Macdonald limit of the lens space index ($p \to 0$) for a free (full-)hypermultiplet is given by
\begin{align} \label{eq:hyperIdx}
\begin{split} 
 \CI^{\textrm{Mac}}_{H, S^3/\IZ_r} (a, m; q, t) & \ = \
 \begin{dcases}
  \frac{1}{(t^{\half} q^{r-m} a; q^r)(t^{\half} q^{m} a^{-1}; q^r)} \,, & \text{when }m \neq 0 \, , \\
  \frac{1}{(t^{\half} a; q^r)(t^{\half} a^{-1}; q^r)}  \,, & \text{when }m=0 \, .  
  \end{dcases}
\end{split}
\end{align}
Here $a$ denotes the flavor fugacity and $m$ denotes the choice of discrete holonomy for the $U(1) \subset SU(2)$ flavor symmetry. We have removed the zero-point energy contributions, which becomes zero as we take $p \to 0$. Throughout this paper, the Macdonald limit $p \to 0$ refers to the limit upon removing the overall zero-point energy contribution of the form $\left(\frac{pq}{t}\right)^\e$, for some $\epsilon \in \mathbb{Q}$. 

Let us first note that the untwisted, refined character for the $\beta-\gamma$ system \eqref{eq:bgUntwisted} agrees with the Macdonald index \eqref{eq:hyperIdx} for $m=0$ upon rescaling 
\bea
q  \ \to \ q^{1/r} \,, \quad \text{ and } \quad t \ = \ qT \,.
\eea
When we turn on the discrete holonomy, the index agrees with the refined character upon rescaling 
\bea
q  \ \to \ q^{1/r} \,, \quad t \ = \ T \,, \quad \text{and} \quad \lambda \ = \ \frac{m}{r} \,.
\eea
The usual (unrefined) character can be obtained by taking $T \to 1$. Notice that the identifications of parameters is different for the twisted ($m\neq 0$) and the untwisted $(m=0)$ cases. In later Sections, we will demonstrate that this is a general phenomenon.

In summary, we find that the (refined) twisted vacuum character of the $\beta-\gamma$ system is identical to the Macdonald/Schur limit of the lens space index:
\beaa
 & \CI^{\textrm{Mac}}_{H, S^3/\IZ_r}(a, m \neq 0; t=q T, q^{1/r}) & = \    & \chi^{\textrm{ref}}_{\CV^{(\frac{m}{r})}[\b\g]}(a; T, q)\,, \quad & (\textrm{when } m \neq 0)\ , \\
 & \CI^{\textrm{Mac}}_{H, S^3/\IZ_r}(a, m=0; t=T, q^{1/r})  & = \ &  \chi^{\textrm{ref}}_{\CV^{(0)}[\b\g]}(a; T, q) \,, \quad & (\textrm{when } m = 0)\ , 
\eeaa
where $\CV^{(\lambda)}[\b\g]$ is the twisted vacuum module (twisted-VOA) generated by the $\beta-\gamma$ ghost system (or symplectic bosons). 


\subsection{The twisted $(b,c)$ ghost system and the free vector multiplet}

The chiral algebra of the free vector multiplet is given by the $(b,c)$ ghost system of conformal weight $(h_b, h_c) = (1,0)$. This is ``essentially" the fermionic version of the $\beta-\gamma$ system discussed above. The chiral algebra is generated by two fields $b(z)$ and $\p c(z)$ that are fermionic 
(\ie~anti-commuting) fields. We remove the zero mode of $c(z)$ in order to match with the free vector multiplet. The OPEs of this system reads
\begin{align}
c(z) b(w) \ = \ b(z) c(w) \ \sim \ \frac{1}{z-w} + \cdots\,,
\end{align}
with nonsingular OPEs amongst each other. Defining the mode expansion
\begin{align}
c(z) \ =\ \sum_{n \in \mathbb{Z}\backslash \{0 \} } c_n z^{-n}\,, \qquad b(z) \ = \ \sum_{n\in \mathbb{Z}} b_n z^{-n-1} \,,
\end{align}
the OPE gives the canonical anti-commutation relations among the modes
\bea
\{c_m, b_n\} \ = \ \delta_{m+n,0} \,.
\eea
The stress-energy tensor is given as
\bea
T(z) \ = \ \sum_{n\in \mathbb{Z}} L_n z^{-n-2} \ = \ :\partial c(z) b(z): \,,
\eea
with normal ordering such that annihilation modes are to the right (with minus sign when fields are interchanged). The vacuum state is defined as 
\bea
c_m \ket{0} \ = \  b_n \ket{0} \ = \ 0 \,, \quad \forall \, m>0\,, \  n \geq 0 \,.
\eea
Thus, we easily see that the (refined) vacuum (super-) character is given by
\begin{align}
 \chi^{\textrm{ref}}_{\CV^{(0)}} (q, T) \ = \ (q; q)(qT; q) \, , 
\end{align}
where the first factor comes from the modes of $\partial c(z)$ and the second factor from the modes of $b(z)$. To get the refined character, we assign the $T$-weights as $w(c) = 0$ and $w(b)=1$~\cite{Song:2016yfd}. 

There is a ghost-number $\mathfrak{u}(1)$ symmetry generated by 
\bea
 J \ = \ :c(z)b(z): \,,
\eea
under which $c$ has charge $+1$ and $b$ has charge $-1$. Now we can define a twist (or boundary conditions)
\bea
c(e^{2\pi \ii}z) \ = \ e^{2\pi \ii \lambda} c(z) \,, \qquad b(e^{2\pi \ii}z) \ = \ e^{ - 2\pi \ii \lambda} b(z) \,,
\eea
where $\lambda \in [0,1)$. This twist does not affect the OPEs, but it does affect the mode expansions, namely
\bea
c(z) \ = \ \sum_{k\in \mathbb{Z}} c_{k-\lambda} z^{-k+\lambda} \,, \qquad b(z) \ = \ \sum_{\ell\in \mathbb{Z}} c_{\ell+\lambda} z^{-\ell-1-\lambda}\,.
\eea
The condition on the vacuum state is the same as before, so that we have
\begin{align}
 c_{k-\lambda} \ket{0} \ = \ b_{\ell+\lambda} \ket{0} \ = \ 0 \,, \quad \textrm{for } k>\lambda \, , \ \ell > -1-\lambda \, .
\end{align}
Now for nonzero $\lambda$, in order to express $J$ and $T$ as a mode expansion we have to include additional terms
\bea
J (z) \ = \ :c(z) b(z): + \frac{\lambda}{z}  \,, \qquad T(z) \ = \ :\partial c(z) b(z): + \frac{\lambda (\lambda-1)}{2z^2}\,,
\eea
where now the normal ordering is such that $c_{n-\lambda}$ and $b_{n-(1-\lambda)}$ with $n>0$ are moved to the right. 
Then the modes contributing to the twisted vacuum character for $\lambda \neq 0$ are
\begin{align}
\begin{split}
 & b_{-1-\lambda}\,,~ b_{-2-\lambda}\,, ~ b_{-3-\lambda}\,, \ldots \,,\\
 & c_{-\lambda}\,, ~c_{-1-\lambda}\,, ~ c_{-2-\lambda}\,, \cdots  \, . 
\end{split}
\end{align}
One can now compute the super-character for a given $\lambda \in [0,1)$~\footnote{For simplicity we remove all the $c/24$-factors and normalize such that the expansion will start as $1+\cdots$.}
\bea
 \chi_{\CV^{(\lambda)}}(a; q) \ = \ \tr \left( (-1)^{F}a^{J_0} q^{L_0} \right) \ = \ ( a q^{\lambda};q )( a^{-1} q^{1-\lambda};q ) \,,
\eea
where the first and the second factor are generated by $b(z)$ and $\p c(z)$ respectively. 
Notice that when $m=0$, or $r=1$, we have to be careful to remove the zero-mode $c_{-\lambda}$, which in turn removes a factor of $(1-a)$, and we get
\begin{align}
 \chi_{\CV^{(\lambda)}}(a; q)  \ = \ (1-a)^{-\delta_{\lambda,0}} ( a q^{\lambda};q ) ( a^{-1} q^{1-\lambda};q ) \,.
\end{align}
One can refine the index by assigning $w(b)=1$ and $w(c)=0$ for the $c(z)$ to obtain
\begin{align}
 \chi^{\textrm{ref}}_{\CV^{(\lambda)}}(a; q, T)  \ = \ (1-a)^{-\delta_{\lambda,0}} ( a q^{\lambda};q ) ( a^{-1} T q^{1-\lambda};q ) \,.
\end{align}

Now, let us compare this result with the lens space index for a free vector multiplet. We shall turn on the flavor fugacity for a single vector multiplet. In the Macdonald limit, only the gauginos contribute to the index. Notice that the two gauginos contributing to the index are charged oppositely under the gauge symmetry. See Table~\ref{table:LetterIndex}, for example. Therefore, the lens index in the Macdonald limit is given as~\footnote{Here we have introduced the notation
\bea
[x]_r & = & \left\{ k \in \{0, \ldots, r-1\} \ \big| \ k \ = \  x \mod r  \right\}\,,
\eea
which we use throughout the paper.}
\begin{align}
\begin{split}
 \CI^{\textrm{Mac}}_{V, S^3/\IZ_r}(a, m) & \ = \ (1-a)^{\delta_{[m]_r, 0}}  (a q^{[m]_r}; q^r)( a^{-1} t q^{[-m]_r}; q^r) \\
  & \ = \ (1-a)^{\delta_{\lambda, 0}}  (a q^{r\lambda}; q^r)(a^{-1} t q^{r(1-\lambda)}; q^r)  \, , 
\end{split}
\end{align}
which agrees with the (refined) character for the $\lambda = \frac{m}{r}$-twisted module we computed above, upon the replacements
\bea
q \ \to \ q^{1/r} \,, \quad \text{ and } \quad t \ \to \ T\,,
\eea
for $\lambda \neq 0$. Here we notice that when $\lambda = 0$, the Macdonald index is identical to the character with a \emph{different} parameter identification $q \to q^{1/r}$ and $t \to q T$. To summarize, 
\begin{align}
&\CI^{\textrm{Mac}}_{V, S^3/\IZ_r}(a, m \neq 0; t=T, q^{1/r}) = \chi^{\textrm{ref}}_{\CV^{(\frac{m}{r})}}(a; T, q) & (\textrm{when } m \neq 0)\, , \\
&\CI^{\textrm{Mac}}_{V, S^3/\IZ_r}(a, m=0; t=q T, q^{1/r}) = \chi^{\textrm{ref}}_{\CV^{(0)}}(a; T, q) & (\textrm{when } m=0)\, .
\end{align}
Akin to the free hypermultiplet, we find that the map of parameters depends on whether we turn on external background holonomies.


\section{Lens space indices of Argyres-Douglas theories} \label{sec:ADindex}

In this Section we outline in some detail how to compute the lens index for Argyres-Douglas theories of type $\left( A_1, A_N \right)$ and $\left( A_1, D_N \right)$. We will use the $\cN=1$ gauge theories that flow to $\cN=2$ Argyres-Douglas theories discovered in~\cite{Maruyoshi:2016tqk,Maruyoshi:2016aim,Agarwal:2016pjo, Agarwal:2017roi, Benvenuti:2017bpg}. See also~\cite{Benvenuti:2017lle, Benvenuti:2017kud, Benvenuti:2017bpg} for the study on the three-dimensional version of the corresponding flows, and see~\cite{Evtikhiev:2017heo} for a study on the conditions for the SUSY enhancement. 

In order to compute the index, we use the $\CN=1$ gauge theory description of the $\CN=2$ fixed point theory. 
This theory has an anomaly-free $U(1)$ global symmetry that can be mixed with the overall R-symmetry (or equivalently two candidate R-symmetries, following the notation of~\cite{Agarwal:2014rua})
\begin{align}
 R_{IR}  \ = \ R_0 +\e \CF  \ = \ \frac{1}{2}(J_+ + J_-) + \frac{\e}{2} (J_+ - J_-) \, . 
\end{align}
The superconformal R-symmetry in the infrared (equivalently the value of $\e$) is determined via $a$-maximization~\cite{Intriligator:2003jj}. 

We will first compute the $\CN=1$ index and then later recast it into an $\CN=2$ index. The $\CN=1$ index is defined as 
\begin{align}
 \CI^{\CN=1} (\fp, \fq; \xi)  \ = \ \tr (-1)^F p^{j_1+j_2 - \frac{R}{2}} q^{j_2-j_1 - \frac{R}{2}} \xi^{\CF} \, , 
\end{align}
where $(j_1, j_2)$ are the Cartans of the Lorentz group and $R$ is the generator of the R-symmetry. We omitted any other flavor fugacities except for the $\CF = \half(J_+ - J_0)$, which is the global $U(1)$ symmetry that mixes with the R-symmetry in our setup. In the end, we recast this into the canonical fugacities for the $\CN=2$ superconformal field theory. The $\CN=2$ index is written as
\begin{align}
 \CI^{\CN=2} (p, q, t)  \ = \ \tr (-1)^F p^{j_1 + j_2 - \frac{r}{2}} q^{j_2 - j_1 - \frac{r}{2}} t^{I-\frac{r}{2}} \, , 
\end{align}
where $I$ is the Cartan of $SU(2)_R$ and $r$ is the generator of $U(1)_r$ symmetry. We can map the $\CN=1$ fugacities to $\CN=2$ fugacities by taking 
\bea
\xi \ \to \ \left(t \left(pq\right)^{-\frac{2}{3}}\right)^\beta \,,
\eea
where $\beta$ has to be fixed separately; it depends on the normalization of $U(1)_\CF$ inside $SU(2)_R \times U(1)_r$.

\subsection{The $(A_1, A_{2N})$ AD theory}\label{secA1A2n}

In order to obtain the lens index of the $(A_1,A_{2N})$ Argyres-Douglas theory, we start with $\cN=1$ $Sp(N)$ SQCD with an adjoint chiral multiplet $\phi$, and two chiral multiplets $q_1, q_2$ in the fundamental representation of $Sp(N)$.~\footnote{Our convention for the symplectic group is such that $Sp(1)=SU(2)$.} We also have $N$ singlets $M_i$, with $i=1, \ldots, N$ which couple to $q_2$, and another $N$ singlets $X_i$ with $i=1, \ldots, N$, which acts as Lagrange multipliers to project out $\tr \phi^{2i}$. The matter content is given in Table~\ref{table:A1A2n}. 
\begin{table}[h]
\centering{\renewcommand{\arraystretch}{1.3}
\begin{tabular}{|c|c|ccc|}
	\hline
	& $Sp(N)$ & $(J_+, J_-)$& $(R_0, \CF)$ & $R_{IR}$ \\
	\hline
	$q_1$ & $N$ & $(1, 0)$ & $(\half, \half)$ & $\frac{6N+8}{6N+9}$ \\
	$q_2$ & $N$ & $(1, -4N-2)$ & $(\frac{-4N-1}{2}, \frac{4N+3}{2})$ & $\frac{2N+6}{6N+9} $\\
	$\phi$ & adj & $(0, 2)$ & $(1, -1)$ & $\frac{2}{6N+9} $ \\
	$M_i$ $(i=1, \ldots, N)$ & $1$ & $(0, 4N+4i+4)$ & $(2N+2i+2, -2N-2i-2)$ & $\frac{4N+4i+4}{6N+9}$ \\
	$X_i$ $(i=1, \ldots, N)$ & $1$ & $(2, 2-4i)$ &$(2-2i, 2i)$ & $\frac{12N+18-4i}{6N+9}$ \\ 
	\hline
\end{tabular}}
\caption{Matter content of the $\CN=1$ gauge theory, which flows to the $(A_1, A_{2N})$ Argyres-Douglas theory. The $(J_+, J_-)$ are two candidate R-charges. The superconformal R-charge $R_{IR} = R_0 + \e \CF$ is given by a linear combination of the two, which is determined via $a$-maximization. From this we obtain $\epsilon =  \frac{6N + 7}{6N + 9}$. The last column denotes the superconformal R-charge at the IR fixed point.}
\label{table:A1A2n}
\end{table}
The superpotential is given as
\begin{align}
 W = \tr(q_1 \phi q_1) + \sum_{i=1}^N M_i \tr (q_2\phi^{2N+1-2i} q_2) + \sum_{i=1}^N X_i \tr \phi^{2i} \, , 
\end{align}
where we take the trace over the gauge indices.~\footnote{This is a simplified version of the original $\CN=1$ gauge theory discussed in~\cite{Maruyoshi:2016tqk,Maruyoshi:2016aim,Agarwal:2016pjo}. Here we removed operators that get decoupled along the RG flow and also introduced Lagrange multiplier fields $X_i$ to take care of the would-be decoupled operators $\tr \phi^{2i}$. In the three-dimensional version of our RG flow this term turns out to be crucial for the supersymmetry enhancement in the IR~\cite{Benvenuti:2017lle, Benvenuti:2017kud, Benvenuti:2017bpg}.} 

We shall now use the $\cN=1$ lens index ingredients as specified in Appendix~\ref{N1indexdefs} to write
\bea \label{eqn:A1A2Nlens}
\cI^{}_{\left( A_1,A_{2N} \right) , \, r} & = & \prod_{i=1}^{N}  \cI_{\chi, r}^{} \left( \xi^{-2(N+i +1)},0,\frac{4N+4i+4}{6N+9} \right) \CI_{\chi, r}^{} \left(\xi^{2i}, 0, \frac{12N+18-4i}{6N+9} \right)  \\
&&\times \sum_{m_\ell=0}^{r-1}\frac{1}{\left| \cW_{m}\right|} \oint \left[\diff \vec{z} \right]_{m} \prod_{\alpha \in \Delta \cup \{0\}} \cI_{\mathrm{V},\, r}^{} \left( z^{\alpha} , \left[\alpha ( m )\right]_{r}\right)
\cI_{\chi,\, r}^{} \left( z^{\alpha} \xi^{-1} , \left[\alpha ( m )\right]_{r}, \frac{2}{6N+9} \right)\nn\\
&&\qquad  \times 
\prod_{\rho \in \mathbf{R}_{\mathrm{f}}} \cI_{\chi,\, r}^{} \left( z^{\rho} \xi^{\tfrac{1}{2}} , \left[\rho ( m )\right]_{r} , \frac{6N+8}{6N+9}\right)
\cI_{\chi,\, r}^{} \left( z^{\rho} \xi^{\tfrac{4N+3}{2}} , \left[\rho ( m )\right]_{r} , \frac{2N+6}{6N+9}\right)\, .\nn
\eea
Here, $\CI_{\chi, r}^{\CN=1}(z, m, R)$ denotes the free $\cN=1$, $L(r,1)$ lens index for a chiral multiplet of R-charge $R$. The variable $z$ is the fugacity for the global/gauge symmetry and $m$ labels the discrete holonomy for the corresponding symmetry. By $\CI_{V, r}^{\CN=1}(z, m)$ we denote the $\CN=1$, $L(r,1)$ vector multiplet contribution, where $m$ denotes the discrete holonomy associated with the gauge group. By $\alpha \in \Delta \cup \{0\}$ we denote the roots of $G=Sp(N)$ including $\mathrm{rk}\left( G \right)$ times the zero-root. Furthermore we have adopted the standard notation that
\bea
z^{\alpha} & = & \prod_{i} z_{i}^{\alpha_i} \,,\quad \text{and} \quad \alpha ( m ) \ = \ \sum_{i} \alpha_{i} m_i \,,
\eea
where $\alpha_i$ are the roots of the gauge group in some basis of the Cartan subalgebra, and similarly for the weights $\rho$ of the representation $\mathbf{R}$. In this instance $\mathbf{R}_{\mathrm{f}}$ denotes the fundamental representation of $Sp(N)$. Also here and in the rest of the paper we denote by $\left[\diff \vec{z} \right]_{m}$ the Haar measure of the corresponding (broken) gauge group, given by
\bea
\left[\diff \vec{z} \right]_{m} & = & \left[\prod_{i=1}^{\mathrm{rk} G} \frac{\diff z}{2\pi \ii}\right] \prod_{\alpha \in \Delta\cup \{0\}} \frac{1}{\left( 1- z^{\alpha} \right)^{\delta_{[\alpha(m)]_r}}} \,.
\eea
Let us also remind the reader that the sum over $m_\ell$ is over all the possible discrete holonomies of $G$, \ie~~$\ell \in \{1, \ldots \mathrm{rk}\left( G \right)\}$. 

In order to write the index in canonical $\CN=2$ fugacities, we replace
\bea
 \xi & \ \rightarrow \ & \left(t \left(p q\right)^{-2/3}\right)^{\frac{1}{2 N+3}} \, , 
\eea
and we arrive at the $\cN=2$ index, which we denote as $\cI^{\cN=2}_{\left( A_1,A_{2n} \right) , \, r} $. Let us remark that in this case there is no freedom to turn any external/flavor discrete holonomies and as we discuss briefly in Section~\ref{sec:A1A2NtwistedCA}, it follows that the lens index in the Macdonald limit will equal the round index upon some rescaling of the fugacities. From the chiral algebra point of view, the reason for that is that the OPEs for the associated minimal models do not allow for any twisting action.

Using Mathematica we can expand the simplest cases in terms of $\ft$, where for convenience we have replaced the fugacities as follows
\bea
p \ \rightarrow \  {\ft}^3 y \,, \qquad q \ \rightarrow \  \frac{{\ft}^3}{y} \,, \qquad t \ \rightarrow \  \frac{{\ft}^4}{v} \,.
\eea
For instance, in the case of the $(A_1, A_2)$ theory, the $L(3,1)$-lens index can be expanded as
\begin{align}
\cI^{}_{\left( A_1,A_{2} \right) , \, 3}(\ft, y, v) & = 
1
+v^{2/5} \ft ^{4/5}
+v^{3/5} \ft ^{6/5}
+v^{6/5} \ft ^{12/5}
+v^{8/5} \ft ^{16/5}
+v^{9/5} \ft ^{18/5}\nn\\
&~~
+\frac{\left(-1+v^3\right) \ft ^{24/5}}{v^{3/5}}
+v^{14/5} \ft^{28/5}
+\left(-1+v^3\right) \ft ^6
-\sqrt[5]{v} \ft ^{32/5} \\
&~~
+v^{3/5} \left(-2+v^3\right) \ft ^{36/5}
-v^{4/5} \ft ^{38/5}
+\left(v+v^4\right) \ft ^8
+\cO\left(\ft^{41/5}\right) \,. \nn
\end{align}
We can revert back to the original $(p,q,t)$-variables and find
\bea
\cI^{}_{\left( A_1,A_{2} \right) , \, 3}(p, q, t) & = & 
1
+\left(p^{2/5}-p^{2/5} t\right) \frac{q^{2/5}}{t^{2/5}}
+\frac{p^{3/5} q^{3/5}}{t^{3/5}}
-p q
+\left(\frac{p^{6/5}}{t^{6/5}}-\frac{p^{6/5}}{\sqrt[5]{t}}\right)q^{6/5}\nn\\
 &&
+\left(\frac{p^{8/5}}{t^{8/5}}-\frac{2 p^{8/5}}{t^{3/5}}\right) q^{8/5}
+\left(\frac{p^{9/5}}{t^{9/5}}-\frac{p^{9/5}}{t^{4/5}}\right)q^{9/5}
+\frac{p^2q^2}{t}
+\frac{p^{12/5} q^{12/5}}{t^{12/5}}\nn\\
 &&
+\frac{p^{14/5} q^{14/5}}{t^{14/5}}
+\frac{p^3 q^3}{t^3}
+\frac{p^{18/5} q^{18/5}}{t^{18/5}}
+\frac{p^4q^4}{t^4}+\cO\left(q^{21/5}\right) \,.\label{expA1A2r3}
\eea
In the Schur limit $p \to 0$ and $t \to q^r$, the index is the same as that of the round sphere index, up to rescaling $q \to q^{1/r}$. We re-iterate that this reflects the fact that there is no twisted module of the chiral algebra associated to the $(A_1, A_{2N})$ theory. Actually we can make this more precise: In the integral~\eqref{eqn:A1A2Nlens} the leading order contribution (in $p$) of the zero-point energy will arise from the sector where $m_{\ell} = 0$. Therefore upon an overall rescaling (such that the index starts with $1+\cdots$) and taking the Macdonald $p \to 0$ limit only the $m_{\ell} = 0$ sector will contribute. It is then easy to see that by taking the limit $q\to q^{1/r}$, the integrand reduces precisely to the round index integrand, and thus we will get the same Macdonald limit independent of the parameter $r$. This is in agreement with chiral algebra considerations as we shall outline in Section~\ref{sec:A1A2NtwistedCA}.

\subsection{The $(A_1, A_{2N-1})$ AD theory}\label{secA1A2n1}

To get the lens index of the $(A_1, A_{2N-1})$ Argyres-Douglas theory, we start with an $\cN=1$ $SU(N)$ gauge theory with two fundamental chiral multiplets and a number of singlets. The matter content and the charges are summarized in Table~\ref{table:A1A2n1}. 
\begin{table}[h]
\centering{\renewcommand{\arraystretch}{1.3}
\begin{tabular}{|c|c|c|ccc|}
	\hline
	& $SU(N)$ & $U(1)_B$ & $(J_+, J_-)$& $(R_0, \CF)$ & $R_{IR}$ \\
	\hline
	$q$ & $N$ & $+1$ & $(1, -2N+1)$ & $(1-N, N) $ & $\frac{N+3}{3N+3}$ \\
	$\tilde{q}$ & $\bar{N}$ & $-1$ & $(1, -2N+1)$ & $(1-N, N)$ & $\frac{N+3}{3N+3}$  \\
	$\phi$ & $\textrm{adj}$ & $0$ & $(0, 2)$ & $(1, -1)$ & $\frac{2}{3N+3}$  \\
	$M_i $ $(i=2, \ldots, N)$ & $1$ &$0$ &$(0, 2N+2i)$ & $(N+i, -N-i)$ & $\frac{2N+2i}{3N+3}$  \\
	$X_i$ $(i=2, \ldots, N)$ & $1$ & $0$ & $(2, 2-2i)$ &  $(2-i, i)$ & $\frac{6N+6-2i}{3N+3}$ \\
	\hline
\end{tabular}}
\caption{Matter contents of the $\CN=1$ gauge theory flowing to the $(A_1, A_{2N-1})$ theory. The $(J_+, J_-)$ are two candidate R-charges. $U(1)_B$ denotes a flavor symmetry. The superconformal R-charge is given by $R_{IR} = R_0 + \e \CF$ where $\epsilon = \frac{3N+1}{3N+3}$. }
\label{table:A1A2n1}
\end{table}
The superpotential of the theory reads
\begin{align}
 W \ = \ \sum_{i=2}^{N} M_i \tr \left(\tilde{q} \phi^{N-i} q \right) + \sum_{i=2}^{N} X_i \tr \phi^{i} \, . 
\end{align}

The lens index of the $\cN=1$ IR theory is then computed as
\begin{align}
\cI^{}_{\left( A_1,A_{2N-1} \right) , \, r} & = 
\prod_{i=2}^{N} \cI_{\chi, \, r}^{} \left( \xi^{-N-i},0,\frac{2N+2i}{3N+3} \right)
\times \prod_{i=2}^{N} \cI_{\chi, \, r}^{} \left( \xi^{i},0,\frac{6N+6-6i}{3N+3} \right) \nn \\
&~~\times \sum_{m_\ell=0}^{r-1} \bigg\{ \frac{1}{\left|\cW_{m}\right|} \oint \left[\diff \vec{z} \right]_{m} 
 \prod_{\alpha\in \Delta\cup \{ 0\}} \cI_{\mathrm{V},\, r}^{} \left( z^{\alpha} ,\left[\alpha( m )\right]_{r} \right) \cI_{\chi,\, r}^{} \left( z^{\alpha} \xi^{-1} ,\left[\alpha( m )\right]_{r} ,\frac{2}{3N+3} \right) \nn\\
&\qquad \times \prod_{\rho \in \mathbf{R}_{\mathrm{f}}} \cI_{\chi,\, r}^{} \left( \left( z^{\rho} a \right)^{\pm} \xi^{N} , \left[\pm ( \rho( m )+m_a )\right]_{r},\frac{N+3}{3N+3} \right) \bigg\}\,.
\end{align}
In the case of the $\left( A_1,A_{2N-1} \right)$ theory, there is a $U(1)$ flavor symmetry (except for the case of $N=2$, where the flavor symmetry is actually enhanced to $SU(2)$), and thus we introduce a continuous fugacity $a$ and a discrete holonomy $m_a \in\{0, \ldots, r-1\}$ for this symmetry. We also used the standard shorthand notation 
\bea
f\left( x^{\pm}, \pm m \right) \ \equiv \ f\left( x , m\right) f( x^{-1} ,-m) \,.
\eea
And finally, as before, by $\Delta\cup \{0\}$ we denote the roots (including zero-roots) of the gauge group $G=SU(N)$ and by $\mathbf{R}_{\mathrm{f}}$ the fundamental weights of $SU(N)$. In order to map to $\cN=2$ fugacities we perform the replacement,
\bea
\tilde \xi & \ \rightarrow \ & \left(t \left(pq\right)^{-2/3}\right)^{\tfrac{1}{N+1}}\,.
\eea

Again using Mathematica we can compute this as a series expansion in $\ft$ and then revert back to $\left( p,q,t \right)$-fugacities. For the example of the $\left( A_1, A_3 \right)$ Argyres-Douglas theory, the $L(3,1)$ lens index with (discrete) $SU(2)$ holonomy $m_a=1$ is given by~\footnote{In order to specify the data for the continuous flavor fugacity $a$ as well as the discrete holonomy $m_a$, we shall denote the $(A_1,A_{2N-1})$ lens index as $\cI^{}_{\left( A_1,A_{2N-1} \right) , \, r} ( a, m_{a} ) $ in the following.}
\begin{align}
\begin{split}
\cI^{}_{\left( A_1,A_{3} \right) , \, 3} ( a, 1 ) &=\left( \frac{p q}{t} \right)^{2/9}\bigg\{1+a^2 t q+\left(\frac{t}{a^2}+a^4 t^2\right) q^2+\left(2 t+a^6 t^3\right) q^3\\
&~~+\left(a^2 t+\frac{t^2}{a^4}+a^2 t^2+a^8 t^4\right) q^4\\
&~~+\left(\frac{t }{a^2}+\frac{t^2}{a^2}+a^4 t^2+a^4
   t^3+a^{10} t^5\right)q^5\\
&~~+\left(2 t+2 t^2+\frac{t^3}{a^6}+a^6 t^3+a^6 t^4+a^{12} t^6\right) q^6\\
&~~+\left(a^2 t+\frac{t^2}{a^4}+3 a^2 t^2+\frac{t^3}{a^4}+a^2 t^3+a^{8} t^4+a^{8}
   t^5\right) q^7\\
&~~+\left(\frac{t}{a^2}+\frac{3 t^2}{a^2}+2 a^4 t^2+\frac{t^3}{a^2}+2 a^4 t^3+\frac{t^4}{a^8}+a^4 t^4+a^{10} t^5\right) q^8\\
&~~+\cO\left(q^{9}\right)\bigg\} + o \left( p^{2/9} \right)\,.
\end{split}
\end{align}
We remove the prefactor coming purely from the zero-point energy and then we take the $p\rightarrow 0$ and $ t \rightarrow 1, \ q\rightarrow q^{1/r} $ limit to get~\footnote{In our ``big and little O"-notation, the terms dubbed $o \left( p^{2/9} \right)$ vanish in that particular limit, \ie~more generally $\lim_{p\rightarrow 0} \frac{o\left( p^{\alpha} \right)}{p^{\alpha}} = 0$, whereas $\lim_{p\rightarrow 0} \frac{\cO\left( p^{\alpha} \right)}{p^{\alpha}} = \mathrm{constant}$, for some $\alpha \in \mathbb{R}$.}
\bea
\cI^{}_{\left( A_1,A_{3} \right) , \, 3} ( a, 1 ) &\rightarrow&
1+a^2 q^{1/3}+\left(\frac{1}{a^2}+a^4\right) q^{2/3}+\left(2+a^6\right) q+\left(\frac{1}{a^4}+2 a^2+a^8\right) q^{4/3}\nn\\
&&
+\left(\frac{2}{a^2}+2 a^4+a^{10}\right)
   q^{5/3}+\left(4+\frac{1}{a^6}+2 a^6+a^{12}\right) q^2\nn\\
&&+\left(\frac{2}{a^4}+5 a^2+2 a^8\right) q^{7/3}
+\cO\left(q^{8/3}\right) \,,
\eea
which precisely agrees with the $\left( \tfrac{1}{3} \right)$-twisted character of the $\widehat{\mathfrak{su}( 2 )}_{-\frac{4}{3}}$ Kac-Moody algebra as we discuss in the next Section. 

\subsection{The $(A_1, D_{2N})$ AD theory}

To obtain the lens index for the $\left( A_1, D_{2N} \right)$ Argyres-Douglas theory, we start with an $\cN=1$ $SU(N)$ infrared theory with two chiral multiplets in the bifundamental and a single one in the adjoint representation of $SU(N)$. The matter content is detailed in Table~\ref{table:A1D2n}. 
\begin{table}[h]
\centering
\begin{tabular}{|c|c|cc|ccc|}
	\hline
	& $SU(N)$ & $U(1)_1$ & $U(1)_2$ & $(J_+, J_-)$ & $R_{IR}$& $\CF$ \\
	\hline
	$q_1$ & $N$ & $1$ & $2N-1$ &$(1, 0)$ & $\frac{3N-1}{3N}$ & $\half$ \\
	$\tilde{q}_1$ & $\bar{N}$ & $-1$ & $-2N+1$ & $(1, 0)$ & $\frac{3N-1}{3N}$ & $\half$ \\	
	$q_2$ & $N$ & $1$ & $-1$ &$(1, -2N+2)$ & $\frac{N+1}{3N}$ & $\frac{2N-1}{2}$ \\
	$\tilde{q}_2$ & $\bar{N}$ & $-1$ & $1$ & $(1, -2N+2)$ & $\frac{N+1}{3N}$ & $\frac{2N-1}{2}$ \\
	$\phi$ & adj & $0$ & $0$ & $(0, 2)$ & $\frac{2}{3N}$ & $-1$ \\
	$M_i$ $(i=2, \ldots, N)$& $1$ & $0$ & $0$ & $(0, 2N+2i-2)$ & $\frac{2N+2i-2}{3N}$ & $1-N-i$ \\
	$X_i$ $(i=2, \ldots, N)$ & $1$ & $0$ & $0$ & $(2, 2-2i)$ & $2 - \frac{2i}{3N}$ & $i$ \\ 
	\hline
\end{tabular}
\caption{Matter contents of the $\CN=1$ gauge theory flowing to the $(A_1, D_{2N})$ theory. Here $U(1)_1$ and $U(1)_2$ denote two flavor symmetries. We also write the superconformal R-charge $R_{IR}=R_0 + \e \CF$ with $\e = \frac{3N-2}{3N}$. }
\label{table:A1D2n}
\end{table}
The superpotential is given as
\begin{align}
 W \ = \ \tr \left(  \tilde{q}_1 \phi q_1 \right) + \sum_{i=2}^N M_i \tr \left( \tilde{q}_2 \phi^{N-i}  q_2 \right) + \sum_{i=2}^N X_i \tr \phi^i \, . 
\end{align} 

Now, the lens index can be computed as
\begin{align}
\cI^{}_{\left( A_1,D_{2N} \right) , \, r} 
\ &= \ \prod_{i=2}^{N} \cI_{\chi , \, r}^{} \left( \xi^{-N-i+1} , 0, \frac{2N+2i-2}{3N} \right) 
{\prod_{i=2}^{N} \cI_{\chi, \, r}^{}\left( \xi^{i},0, 2 - \frac{2i}{3N} \right)} \nn \\
&~ \times \sum_{m_\ell=0}^{r-1}\Bigg\{\frac{1}{\left|\cW_{m} \right|} \oint \left[\diff \vec{z} \right]_{m} 
\prod_{\alpha \in \Delta \cup \{0\}} \cI_{\mathrm{V}, \, r}^{}\left( z^{\alpha}, \left[\alpha( m )\right]_{r} \right) \cI_{\chi, \, r}^{}\left( z^{\alpha} \xi^{-1}, \left[\alpha( m )\right]_{r}, \frac{2}{3N} \right)\nn\\
&~\times \prod_{\rho \in \mathbf{R}_{\mathrm{f}}}
\bigg[
\cI_{\chi,\, r}^{}\left( \left( z^{\rho} a_{1} a_{2}^{-1} \right)^{\pm} \xi^{\tfrac{2N-1}{2}} , \left[\pm \left( \rho ( m) + m^{(1)}_{a}-m^{(2)}_{a} \right)\right]_{r}, \frac{N+1}{3N} \right)\nn\\
&~\times
 \cI_{\chi,\, r}^{}\left( \left( z^{\rho} a_{1} a_{2}^{2n-1} \right)^{\pm} \xi^{\tfrac{1}{2}} , \left[\pm \left( \rho (m) + m_{a}^{(1)}+(2n-1) m_{a}^{(2)} \right)\right]_{r}, \frac{3N-1}{3N} \right)\bigg]\Bigg\}\,,\nn\\
 \label{IA1D2N}
\end{align}
where we have used the same notation as introduced previously. The theory has a $U(1)_{1} \times U(1)_{2}$ symmetry and thus we have introduced corresponding continuous fugacities $a_i$ and discrete flavor holonomies $m_{a}^{(i)}$, for $i=1,2$. In the infrared, the flavor symmetry is enhanced to $SU(2) \times U(1)$, for $N>2$ and to $SU(3)$, for $N=2$. In order to write it as a canonical $\CN=2$ index, we perform the redefinition 
\bea
\xi \ \rightarrow \ \left(t \left(p q\right)^{-2/3}\right)^{\frac{1}{N}} \,.
\eea

In this paper, we pay particular attention to the case of the $\left( A_1, D_4 \right)$ Argyres-Douglas theory, where the associated two-dimensional chiral algebra is given by the $\widehat{\mathfrak{su}( 3 )}_{-\frac{3}{2}}$ affine Kac-Moody algebra. This is thus far the only available $\CN=2$ SCFT with $SU(3)$ flavor symmetry of which we can compute the lens space index. This theory provides a non-trivial test case for our conjecture. 

\subsubsection{The $(A_1, D_{4})$ AD theory}\label{secA1D4}

The $\left( A_1, D_4 \right)$ Argyres-Douglas theory has $SU(3)$ as an enhanced flavor symmetry. In order to observe this enhancement in the index, we perform the following replacements of the $U(1)_1 \times U(1)_2$ flavor fugacities
\bea\label{Eqn:redefA1D4a} 
a_{1} \ \rightarrow \ a_{2}^{3/4} \,, \quad a_{2} \ \rightarrow \ a_{1}^{1/2} a_{2}^{1/4} \,,
\eea
and discrete holonomies $m_a^{(i)}$
\bea\label{Eqn:redefA1D4m}
m_a^{(1)} \ \rightarrow \ \frac{3}{4} m_a^{(2)} \,, \quad m^{(2)}_{a} \ \rightarrow \ \frac{1}{2} m_{a}^{(1)} + \frac{1}{4} m_{a}^{(2)} \,.
\eea
Notice that given the integrand in~\eqref{IA1D2N}, the new flavor holonomies only make sense if their difference is an even number, due to taking the whole expression modulo $r$. As we shall see further below, matching the corresponding twisting of characters of $\widehat{\mathfrak{su}( 3 )}_{-\frac{3}{2}}$ to choices of $m_{a}^{(i)}$ seems somewhat arbitrary because of precisely the above redefinition of $m_a^{(i)}$. 

Now we can compute the lens index of the $\left( A_1, D_{4} \right)$ theory for a variety of examples. For instance in the case of the $L(3,1)$ lens index with a choice of $(m_a^{(1)} ,m^{(2)}_a) = (3,1)$, we can expand the index as~\footnote{As before, in order to specify the data for the continuous flavor fugacities $a_{i}$ as well as the discrete holonomies $m_a^{(i)}$, we shall denote the $(A_1,D_{2N})$ lens index as $\cI^{}_{\left( A_1,D_{2N} \right) , \, r} ( (a_{1}, m_{a}^{(1)}),(a_{2}, m_{a}^{(2)}) ) $ in the following.}
\bea
&& \cI^{\textrm{Mac}}_{\left( A_1,D_{4} \right) , \, 3} \left( ( a_1,3 ),(a_2,1) \right) \nn \\
&&  \ = \ 
1
+ \left[\frac{1}{{a_1}^2 {a_2}}+{a_1} {a_2}^2+\frac{{a_1}}{{a_2}}\right] t q\nn\\
&&~~
+ \bigg[t \left({a_1}^2 {a_2}+\frac{1}{{a_1} {a_2}^2}+\frac{{a_2}}{{a_1}}\right)
+t^2 \left(\frac{1}{{a_1}^4 {a_2}^2}+{a_1}^2 {a_2}^4+\frac{{a_1}^2}{{a_2}^2}\right)\bigg]q^2\nn\\
&&~~
+ \bigg[3 t+t^2 \left({a_1}^3 {a_2}^3+\frac{1}{{a_1}^3 {a_2}^3}+{a_1}^3+\frac{1}{{a_1}^3}+{a_2}^3+\frac{1}{{a_2}^3}\right)
+t^3 \left(\frac{1}{{a_1}^6 {a_2}^3}+{a_1}^3 {a_2}^6+\frac{{a_1}^3}{{a_2}^3}\right)\bigg]q^3 \nn\\
&&~~
+\bigg[
t \left(\frac{1}{{a_1}^2 {a_2}}+{a_1} {a_2}^2+\frac{{a_1}}{{a_2}}\right)
+t^2 \left({a_1}^4 {a_2}^2+\frac{1}{{a_1}^2 {a_2}^4}+\frac{{a_2}^2}{{a_1}^2}+\frac{2}{{a_1}^2 {a_2}}+2 {a_1} {a_2}^2+\frac{2 {a_1}}{{a_2}}\right)\nn\\
&&~~\quad
+t^3 \left(\frac{1}{{a_1}^5 {a_2}^4}+\frac{1}{{a_1}^5 {a_2}}+{a_1}^4 {a_2}^5+\frac{{a_1}^4}{{a_2}}+{a_1} {a_2}^5+\frac{{a_1}}{{a_2}^4}\right)
+t^4 \left(\frac{1}{{a_1}^8 {a_2}^4}+{a_1}^4 {a_2}^8+\frac{{a_1}^4}{{a_2}^4}\right)
\bigg]q^4 \nn\\
&&~~
+\bigg[
t \left({a_1}^2 {a_2}+\frac{1}{{a_1} {a_2}^2}+\frac{{a_2}}{{a_1}}\right)
+t^2 \left(\frac{1}{{a_1}^4 {a_2}^2}+{a_1}^2 {a_2}^4+\frac{{a_1}^2}{{a_2}^2}+3 {a_1}^2 {a_2}+\frac{3}{{a_1} {a_2}^2}+\frac{3 {a_2}}{{a_1}}\right)\nn\\
&&~~\quad
+t^3 \left({a_1}^5 {a_2}^4+{a_1}^5 {a_2}+\frac{1}{{a_1}^4 {a_2}^5}+\frac{2}{{a_1}^4 {a_2}^2}+\frac{{a_2}}{{a_1}^4}+2 {a_1}^2 {a_2}^4+\frac{2 {a_1}^2}{{a_2}^2}+\frac{1}{{a_1} {a_2}^5}+\frac{{a_2}^4}{{a_1}}\right)\nn\\
&&~~\quad
+t^4 \left(\frac{1}{{a_1}^7 {a_2}^5}+\frac{1}{{a_1}^7 {a_2}^2}+{a_1}^5 {a_2}^7+\frac{{a_1}^5}{{a_2}^2}+{a_1}^2 {a_2}^7+\frac{{a_1}^2}{{a_2}^5}\right)\nn\\
&&~~\quad
+t^5 \left(\frac{1}{{a_1}^{10} {a_2}^5}+{a_1}^5 {a_2}^{10}+\frac{{a_1}^5}{{a_2}^5}\right)
\bigg]q^5 
+\cO(q^{6})\,,
\eea
where again we have removed the prefactor coming from the zero-point energy prior to taking the $p\rightarrow 0$ limit. In the $t \rightarrow 1$ limit, this precisely matches the $\left( \frac{1}{3},\frac{1}{3} \right)$-twisted character of $\widehat{\mathfrak{su}( 3 )}_{-\frac{3}{2}}$ for some given vacuum. For the convenience of the reader, and since it is one of the prominent checks of our conjecture relating the lens index to twisted characters, we shall provide several examples of the lens index of the $\left( A_1,D_4 \right)$ theory in Section~\ref{A1D4examples}.

\subsection{The $(A_1, D_{2N+1})$ AD theory}\label{secA1D2n1}

The $(A_1, D_{2N+1})$ Argyres-Douglas theory can be obtained by the renormalization group flow from an $\cN=1$ theory with $Sp(N)$ gauge group, with matter given by two chiral multiplets in the fundamental and another chiral multiplet in the adjoint representation of $Sp(N)$. On top of this, we add a number of singlets. The matter content is given as in Table~\ref{table:A1D2n1}. 
\begin{table}[h]
\centering{\renewcommand{\arraystretch}{1.3}
\begin{tabular}{|c|c|c|ccc|}
	\hline
	& $Sp(N)$ & $SO(3)$ & $(J_+, J_-)$ & $R_{IR}$ & $\CF$  \\
	\hline
	$q_1$ & $2N$ & $3$ & $(1, 0)$ & $\frac{6N+2}{6N+3}$ & $\half$ \\
	$q_2$ & $2N$ & $1$ & $(1, -4N)$  & $\frac{2N+2}{6N+3} $ & $\frac{4N+1}{2}$ \\
	$\phi$ & adj & $1$ & $(0, 2)$ & $\frac{2}{6N+3} $ & $-1$ \\
	$M_i$ $(i=1, \ldots, N)$ & $1$ & $1$ & $(0, 4N+4i)$ & $\frac{4N+4i}{6N+3}$ & $-2N-2i$ \\
	$X_i$ $(i=1, \ldots, N)$ & $1$ & $1$ & $(2, 2-4i)$ & $\frac{12N+6-4i}{6N+3}$ & $2i$  \\ 
	\hline
\end{tabular}}
\caption{Matter content of the $\CN=1$ gauge theory flowing to the $(A_1, D_{2N+1})$ theory. The superconformal R-charge is given by $R_{IR} = R_0 + \e \CF$ in terms of the IR data, with $\e$ determined via $a$-maximization. By doing so, we obtain $\epsilon =  \frac{6N +1}{6N + 3}$. The last column denotes the superconformal R-charge at the IR fixed point.}
\label{table:A1D2n1}
\end{table}
Furthermore, the superpotential reads
\begin{align}
 W \ = \ \tr q_1 \phi q_1 + \sum_{i=1}^N M_i \tr q_2 \phi^{2N+1-2i} q_2 + \sum_{i=1}^N X_i \tr \phi^{2i} \, . 
\end{align}
Notice that we remove all the would-be Coulomb branch operators $\tr \phi^{2i}$ from the chiral ring. In the IR, the $M_i$'s become the operators parametrizing the $\CN=2$ Coulomb branch moduli space. 

Now, the lens index is computed as
\bea
&& \cI^{}_{\left( A_1,D_{2N+1} \right) , \, r} ( \{a_i, m^{(i)}_{a}\}  )
\ = \ \prod_{i=1}^{N} {\cI_{\chi , \, r}^{} \left( \xi^{-2N-2i} , 0 , \frac{4N+4i}{6N+3} \right)}
{\cI_{\chi , \, r}^{}\left( \xi^{2i}, 0, \frac{12N+6-4i}{6N+3} \right)} \nn\\
&&\times \sum_{m_\ell=0}^{r-1} \Bigg\{\frac{1}{\left| \cW_{m} \right|} \oint \left[\diff z\right]_m
\prod_{\alpha \in \Delta\cup\{0\}}\cI_{\mathrm{V}, \, r}^{}\left( z^{\alpha}, \left[\alpha(m)\right]_r \right) \cI_{\chi , \, r}^{}\left( z^{\alpha}\xi^{-1} , \left[\alpha(m)\right]_r , \frac{2}{6N+3} \right)\nn\\
&&\qquad\qquad \qquad \times \prod_{\rho \in \mathbf{R}_{\mathrm{f}}} 
\cI_{\chi , \, r}^{} \left[\prod_{\nu \in \mathbf{V} }\left( z^{\rho} a^{\nu} \xi^{\tfrac{1}{2}} , \left[\rho( m ) + \nu ( m_a )\right]_{r} , \frac{6N+2}{6N+3} \right)\right]\nn\\
&&\qquad\qquad \qquad \quad \qquad \times\cI_{\chi , \, r}^{} \left( z^{\rho} \xi^{\tfrac{4N+1}{2}} , \left[\rho( m )\right]_{r} , \frac{2N+2}{6N+3} \right)
\Bigg\} \,.
\eea
Here $\nu \in \mathbf{V}$ denotes the product over all weights 
of the vector representation of $SO(3)$, which is the flavor symmetry group of the $\left( A_1,D_{2N+1} \right)$ theory. In order to arrive at the $\cN=2$ lens index of the Argyres-Douglas theory, we make the replacement 
\bea
\xi \ \rightarrow \ \left( t \left(p q\right)^{2/3}\right)^{\frac{1}{2 N+1}}\,.
\eea

Again we can compute the index as an expansion in the fugacities $\left( p,q,t \right)$. For matching with twisted characters of the associated chiral algebra, we are mainly interested in the case $\left( A_1,D_{3} \right)$ (which is equivalent to $\left( A_1,A_3 \right)$). For computational reasons however, we focus on the computation of the $\left( A_1,A_3 \right)$ lens index. Nevertheless theories with $N\geq 2$ are interesting since the protected chiral algebra is given by the Kac-Moody algebras $\widehat{\mathfrak{su}( 2 )}_{k}$ with affine level $k=- \frac{4N}{2N+1}$. We have checked some cases of the $\left( A_1,D_5 \right)$ lens index, and matched it to twisted characters of the corresponding chiral algebra. Even though we shall not explicitly explore them more in this paper, let us provide an example of the lens index of $\left( A_1, D_5 \right)$; namely we take 
\bea
r=3 \,, \quad \text{and} \quad ( m_a^{(1)},  m_a^{(2)} )=\left( 1,2 \right) \,.
\eea
Upon removing the zero-point energy, and taking the Macdonald limit $(p\to 0)$, we find the following index
\begin{align} \label{eq:A1D5idx}
\begin{split}
&\CI_{(A_1, D_5), 3}^{\rm Mac} \left( (a , 1),(a^{-1},2) \right) 
= \left( 1+t+t^2+\cdots \right) \\
& \qquad \qquad +\frac{1}{a^2}\left(t+t^2+\cdots \right) q+a^2 \left( t + \cdots \right) q^2  
+(2t + \cdots) q^3 + \cO\left( q^{4} \right) \,.
\end{split}
\end{align}
It turns out that it is computationally extremely labor intensive to go to higher order than this, but we already see a pattern emerging; in particular we see that if we fix a certain vacuum in which $J^{0}_{0} \ket{\rm vac}  \neq 0$ this is precisely the form a twisted refined character is expected to have. The fact that we have to pick a particular vacuum is not particularly surprising, as we shall see in the following.


\section{Twisted chiral algebras for Argyres-Douglas theories }\label{SecAD}

In this Section we shall first describe in some detail the twisted versions of the chiral algebras associated to some Argyres-Douglas theories and then explicitly compare them to limits of the lens index of the corresponding four-dimensional superconformal field theories, as computed in the previous Section.

\subsection{The $\left( A_1,A_{2N} \right)$ theory and its twisted chiral algebra} \label{sec:A1A2NtwistedCA}

The chiral algebra corresponding to the $\left( A_1,A_{2N} \right)$ Argyres-Douglas theory is given by the $\left( 2,2N+3 \right)$ Virasoro minimal model. In this case we do not have any freedom of introducing twisted boundary conditions. The only field in the chiral algebra is the stress-energy tensor, which cannot have any twisting action, without modifying its OPE. Similarly we have seen that in the lens index there was no way to include any nontrivial discrete holonomies, because there is no global flavor symmetry. 

Given these reasons, we expect that after stripping off the overall leading order $\left( \frac{p q}{t} \right)^{\e}$ - prefactor (for some $\e \in \mathbb{Q}$), taking the $p\rightarrow 0$ limit, and then rescaling $q\rightarrow q^{1/r}$, we end up with the usual untwisted vacuum character of the Virasoro minimal models from the lens space index. One can explicitly check this in a variety of examples. For instance, taking the $p \rightarrow 0 $ limit in~\eqref{expA1A2r3}, we end up precisely with the usual round Schur index/Minimal model vacuum character of the $\left( 2,5 \right)$ Virasoro minimal model.

It might look like a somewhat trivial check, but it is worth pointing out that prior to taking the $p\rightarrow 0$ limit, the lens index is vastly different from the round index for the $(A_1,A_{2N})$ Argyres-Douglas theories.

\subsection{The $\left( A_1,D_{2N+1} \right)$ theory and its twisted chiral algebra}

In the case of the $\left( A_1,D_{2N+1} \right)$ Argyres-Douglas theories, the dual chiral algebra is given by the Kac-Moody algebras $\widehat{\mathfrak{su}(2)}_k$ of level
\bea
k & \ = \ & - \frac{4 N}{2N+1} \,.
\eea
In this Section we shall first explicitly describe the possible twistings of $\widehat{\mathfrak{su}( 2 )}_k$. Subsequently, we shall explicitly compare the twisted characters of such twisted affine algebras to the lens index of the $\left( A_1,D_{3} \right)\equiv\left( A_1,A_{3} \right)$ Argyres-Douglas theory.

\subsubsection{The twisted $\widehat{\mathfrak{su}( 2 )}_{k}$ chiral algebra}

The $\widehat{\mathfrak{su}( 2 )}_k$ affine Kac-Moody algebra can be written in terms of three currents $J^{a}\,,$ $a=1,2,3$ satisfying the following OPEs
\bea
J^{a} \left( z \right)J^{b} \left( z \right) & \sim & \frac{k \delta_{ab}}{\left( z-w \right)^{2}} + \sum_c \ii \epsilon_{abc} \frac{J^{c}(w)}{z-w} + \cdots \,.
\eea
In order to obtain raising/lowering operators, we define
\bea
J^{1} & = & \frac{1}{\sqrt{2}} \left( J^{+} + J^{-} \right) \,, \qquad J^{2}  \ = \ - \frac{\ii}{\sqrt{2}} \left( J^{+} - J^{-} \right) \,, \qquad J^{0} \ = \  J^{3} \,,
\eea
such that we have the following system of OPEs
\bea
 J^{0} \left( z \right)J^{\pm}\left( w \right) & \sim & \pm \frac{J^{\pm}(w)}{\left( z-w \right)} + \cdots \,,\\
 J^{+} \left( z \right)J^{-}\left( w \right) & \sim & \frac{k}{\left( z-w \right)^{2}} + \frac{2 J^{0}(w)}{\left( z-w \right)} + \cdots \,,\\
 J^{0} \left( z \right)J^{0}\left( w \right) & \sim &  \frac{k}{2 \left( z-w \right)^{2}} + \cdots \,.
\eea
Now let us introduce twisted boundary conditions for the fields as follows
\bea
J^{0} \left( z \right) & \rightarrow & J^{0} ( e^{2\pi \ii } z) \ = \ J^{0} \left( z \right)\,,\\
J^{\pm} \left( z \right) & \rightarrow & J^{\pm} ( e^{2\pi \ii }z ) \ = \ e^{\pm 2\pi \ii \lambda} J^{\pm}\left( z \right)\,.
\eea
It is easy to see that such a twisting action with arbitrary $\lambda \in [0,1)$ leaves the above OPEs invariant. The mode expansions consistent with the twisting action are then 
\bea
J^{0} \left( z \right) & = & \sum_{k \in \mathbb{Z}} J^{0}_{k} z^{-k-1}\,,\\
J^{+} \left( z \right) & = & \sum_{k \in \mathbb{Z}} J^{+}_{k-\lambda} z^{-k-1+\lambda}\,,\\
J^{-} \left( z \right) & = & \sum_{k \in \mathbb{Z}} J^{-}_{k+\lambda} z^{-k-1-\lambda}\,.
\eea
It is now a simple exercise to check that the modes satisfy the following commutation relations
\begin{align}
\left[J^{0}_m, J_{n\mp \lambda}^{\pm}\right] & = \pm J^{\pm}_{m + n \mp \lambda} \,,\\
\left[J^{+}_{m-\lambda}, J_{n + \lambda}^{-}\right] & = 2 J_{m+n}^{0} + k \left( m -\lambda\right) \delta_{m+n,0} \,,\\
\left[J^{0}_m, J_{n}^{0}\right] & = \frac{km}{2} \delta_{m+n,0} \,.
\end{align}
The energy-momentum tensor
\bea
T\left( z \right) & = & \sum_{n \in \mathbb{Z}} L_{n} z^{-n-2} \,,
\eea
can be given in terms of the affine current generators via the Sugawara construction.  Being careful of keeping track of the zero-modes as well as the factors of $\lambda$, we find for $L_0$
\bea
L_0 & = & \frac{1}{k+2} \Bigg(\sum_{m \in \IZ_{>0}} 2 J^{0}_{-m} J^{0}_{m} + \left( J^{+}_{-m-\lambda} J^{-}_{m+\lambda} +J^{-}_{-m+\lambda} J^{+}_{m-\lambda} \right)  \nn\\
&&\qquad \qquad
+  J_{0}^{0} J_{0}^{0} +  J^{+}_{-\lambda} J^{-}_{\lambda}+ \left( 1- 2\lambda \right) J_0^{0} - \left(1-2\lambda+h_0 \right)h_0\Bigg)\,,
\eea
where $h_0$ labels the ``vacuum", \ie~in the vacuum representation
\bea
J_{0}^{0} \ket{ h_0 } \ = \ h_0 \ket{h_0} \,, \quad J_{\nu}^{\pm} \ket{h_0} \ = \ 0 \,, \quad \nu \geq 0\,.
\eea
Given $L_0$, the commutation relations read
\bea
\left[L_0,J^{+}_{-\ell - \lambda}\right]  & = & \frac{\ell+\lambda}{2} J^{+}_{-\ell - \lambda}\,,\\
\left[L_0,J^{-}_{-\ell + \lambda}\right]  & = & \frac{\ell-\lambda}{2} J^{-}_{-\ell + \lambda}\,.
\eea
It is now straightforward to write down the twisted character for this theory as an expansion by counting states and keeping track of null vectors.

\subsubsection{Twisted character from the lens index}

Let us now explicitly compare the twisted refined character of $\widehat{\mathfrak{su}(2)}_{-\frac{4}{3}}$ with the Macdonald limit of the lens index $\cI_{( A_1,A_3 ), \, r}^{}$ of the $\left( A_1,D_3 \right)=\left( A_1,A_3 \right)$ Argyres-Douglas theory, computed in the previous Sections~\ref{secA1A2n1} and~\ref{secA1D2n1}. 

Let us write down the definition of refined character \cite{Song:2016yfd} for the affine Lie algebra $\widehat{\mathfrak{g}}_{k}$ algebra. The definition for the current case almost goes through as in section \ref{sec:refinedchar}, except that we need to take into account the relation between the Virasoro generators $L_n$ and the current generators $J_n^{i}$. This is due to the fact that we would like to assign the same $T$-weights for the $L_n$ and $J_n$, $w(L)=w(J)=1$, but there is a non-linear Sugawara relation between $L$'s and $J$'s. 
Therefore, we consider the filtration $\CV_0 \subset \CV_1 \subset \CV_2 \subset \ldots$\footnote{There is no half-integral $T$-weight generators in this chiral algebra.}, where
\begin{align} \label{eq:su2filter}
 \CV_k = \mathrm{span} \left\{ X^{(i_1)}_{-\a_1} \cdots X^{(i_m)}_{-\a_m} \ket{\Omega} : i_1 + \cdots + i_m \le k \right\} \Big/ \{\textrm{null states \& relations} \} \ , 
\end{align}
with $X^{(i)} \in \{J^i, L \} $. Then we construct the associated graded vector space $V_{gr} = \bigoplus_i V_i = \bigoplus_i (\CV_i / \CV_{i-1}) $. As before, each $V_i$ can be further decomposed into the eigenspaces of $L_0$ as $V_i = \bigoplus_h V_i^{(h)}$, so that we define the refined character as
\begin{align} \label{eq:su2refchar}
 \chi^{\textrm{ref}}_{\CV}(q, T) \ = \ \sum_{i\ge0} \sum_h \chi(V_h^{(i)}) q^{h} T^i \, , 
\end{align}
where $\chi(V_h^{(i)})$ denotes the (finite) $\mathfrak{g}$-character.\footnote{We removed $(-1)^F$ since we do not have any fermionic generators in this chiral algebra.}

The refined character for the twisted module can be computed by first enumerating the states in the Verma module generated by negative modes of $J^{+}, J^-$ and $J^0$, and then removing the null states and their descendants. The refined character for the Verma module can be easily computed (see also section 3 of \cite{Song:2016yfd}) to give
\begin{align} \label{eq:su2verma1}
\begin{split}
 \chi^{\textrm{ref}}_{\CV^{(\lambda)}_{\textrm{Verma}}} (a; q, T) 
 & \ = \  \textrm{PE} \left[\frac{1}{1-q}\left(qT + q^{1-\lambda}T a^2 + q^{\lambda}T a^{-2} \right) + \frac{q T}{1-q} - \frac{q T^2}{1-q} \right]_{q, T, a} \\
 & \ = \  \prod_{n = 0}^\infty \frac{(1-q^{n+2}T^2)}{(1-q^{n+1} T)(1-q^{1-\lambda+n} T a^2)(1-q^{\lambda+n} T a^{-2})(1-q^{n+2}T)} \, , 
\end{split}
\end{align}
where we have removed the overall shift that comes from $c/24$ or $h_0$. The plethystic exponential is defined as 
\begin{align}
\mathrm{PE}\left[ f \left( \{x_i\} \right) \right]_{\{ x_i\}} \  = \ \exp\left( \sum_{n=1}^{\infty} \frac{f \left(\{x_i^{n}\}\right)}{n}\right) \,.
\end{align}
The first term in the PE comes from the twisted $\widehat{\mathfrak{su}(2)}$ generators, and the last two terms come from the Virasoro generators $L_{-n}$ and the Sugawara relation of the form $L_{n} \sim \sum_m J^a_{n-m} J^a_m$. The $\widehat{\mathfrak{su}(2)}$ and the Virasoro generators contribute more or less independently in the refined character, up to the relation among them given by the Sugawara construction. This relation is responsible for the negative sign of the last term of \eqref{eq:su2verma1} inside the plethystic exponential. Let us notice some differences here as compared to the untwisted vacuum module. One thing to notice here is that the last two terms inside the PE of \eqref{eq:su2verma1} start at order $q$ instead of order $q^2$, differ from the untwisted case. In the case of untwisted vacuum module, the $L_{-1} \ket{0}$ state is null. But for the twisted module, $L_{-1} \ket{0}$ state is not null in general. And we also have a Sugawara relation even at the level $1$ of the form $L_{-1} \sim J^-_{\lambda} J^+_{1-\lambda} $ with $\lambda > 0$. When there is no null state in the twisted module $\CV^{(\lambda)}_{h_0}$ with the highest-weight $h_0$, this is the refined character for the non-degenerate module. Here, we find that there are null states for the $\widehat{\mathfrak{su}(2)}_{-\frac{4}{3}}$-module if we choose $h_0$ appropriately. Moreover, we claim that it is identical to (the Macdonald limit of) the lens index of the $(A_1, D_{3})$ Argyres-Douglas theory upon a suitable map of parameters. 

Let us first summarize the map and then discuss a number of examples in detail. In order to match with the character, we first strip off the factor
\begin{align}\label{Eqn:I0SU2}
\cI^{(0)}_{r, \, \lambda} \ = \ \left( \frac{p q}{t} \right)^{\frac{r k \left( 1-\lambda \right)\lambda}{4}}\,,
\end{align}
from the lens index as before. Now, the twisting parameter of the chiral algebra is given in terms of the discrete holonomy  as
\begin{align}\label{su2lambdavsm}
\lambda \ = \  \frac{[2m]_r}{r} \,.
\end{align}
We then take the $p\rightarrow 0$ Macdonald limit\footnote{Recall that $\CI^{(0)}(p, q, t)$ is the zero-point energy of the $(A_1,A_3)$ Argyres-Douglas theory. In particular it is precisely given by $\cI^{(0)}_{r, \, \lambda}$ in equation~\eqref{Eqn:I0SU2} upon identification of $\lambda$ as given in~\eqref{su2lambdavsm}.} 
\bea
\CI^{\textrm{Mac}} (q, t) \ =  \ \lim_{p \to 0} \, \left[\CI(p, q,t)/\CI^{(0)}(p, q, t) \right] \,,
\eea
 rescale $q\rightarrow q^{1/r}$, and finally either put $t\rightarrow q T$, if $[2m]_r=0$, or $t\rightarrow T$, otherwise. The former limit will always lead to the round index in the ``Schur" limit ($t\to q$) and thus we shall name it the \emph{round limit}. Together we get the following identification
\begin{align}
 \chi^{\textrm{ref}}_{\CV^{(\lambda)}_{h_0}} (a; q, T) = 
 \begin{dcases}
  \CI_{(A_1, A_3), r}^{\textrm{Mac}} (a, m \neq 0; q \to q^{1/r}, t \to T) \, , & \quad (\lambda \neq 0) \\
  \CI_{(A_1, A_3), r}^{\textrm{Mac}} (a, m = 0; q \to q^{1/r}, t \to q T) \, , & \quad (\lambda = 0) \\  
 \end{dcases}
\end{align}
where $\chi_{\CV^{(\lambda)}_{h_0}}$ denotes the $\lambda$-twisted $\widehat{\mathfrak{su}(2)}_{-\frac{4}{3}}$ character for a shifted ``vacuum"
\bea
\ket{ 0} & \ \to \ & \ket{h_0 } \ = \ \ket{ \frac{2\lambda-1}{3} }\,.
\eea
When we take $T \to 1$, we recover the twisted character of the corresponding module. We demonstrate this correspondence in detail for a few cases now.

\paragraph{Lens index for $L(4,1)$.}

Let us consider the lens index for $r=4$ and with the external discrete holonomy $m =  1$. The Macdonald limit of the index (after removing the zero point piece) takes the form
\begin{align}
\begin{split}
\cI_{(A_1,D_3), \, r=4}^{\textrm{Mac}} ( a,m =1 )
&=
1
+q^2 \left(\frac{t}{a^2}+a^2 t\right)
+q^4 \left(2 t+\frac{t^2}{a^4}+a^4 t^2\right)\\
& \quad
+q^6 \left(\frac{t}{a^2}+a^2 t+\frac{t^2}{a^2}+a^2 t^2+\frac{t^3}{a^6}+a^6 t^3\right)\\
& \quad
+q^8 \left(a^8 t^4+\frac{t^4}{a^8}+a^4 t^3+\frac{t^3}{a^4}+a^4 t^2+\frac{t^2}{a^4}+2 t^2+2 t\right)\\
& \quad
+q^{10} \left(3 a^2 t^2+\frac{3 t^2}{a^2}+a^2 t+\frac{t}{a^2}\right)\\
& \quad +\cO\left( q^{12} \right) \,.
\end{split}
\end{align}
On the other hand, the refined character for the Verma module for $\lambda = \half$ is given as
\begin{align} \label{eq:su2verma}
\begin{split}
\chi^{\textrm{ref}}_{\CV^{(1/2)}_{\textrm{Verma}}}  &= 
1+q^{1/2} \left( a^2 + \frac{1}{a^2} \right) T + q \left[ 2T+ \left(a^4+\frac{1}{a^4}\right) T^2\right]  \\
&\quad +q^{3/2} \left[ \left(a^2+\frac{1}{a^2}\right) T +\left(2a^2+\frac{2}{a^2}\right) T^2+\left(\frac{1}{a^6}+a^6 \right) T^3\right]  \\
&\quad +q^2 \left[2T+\left(a^4+\frac{1}{a^4}+4\right) T^2 + \left(2a^4+\frac{2}{a^4}\right) T^3+\left(a^8+\frac{1}{a^8}\right) T^4\right]  \\
& \quad + \cO(q^{5/2}) \, . 
\end{split}
\end{align}

Now, it turns out that taking the $t \rightarrow T$ limit and rescaling $q\rightarrow q^{1/4}$, this precisely matches with the $\left( \lambda , 1-\lambda \right)=\left( \frac{2}{4}, \frac{2}{4} \right)$-twisted refined character of $\widehat{\mathfrak{su}(2)}_{-\frac{4}{3}}$ with vacuum
\bea
 \big| h_0 \big\rangle  \ = \  \big| 0 \big\rangle \,.
\eea
The first null states of the twisted chiral algebra appear at order $q^{3/2} T^2$, given by
\begin{align} \label{eq:su2null1}
\left(\frac{2}{9} J^+_{-\frac{3}{2}} - \frac{2}{3} J^+_{-\half} J^0_{-1} + J^+_{-\half} J^+_{-\half} J^-_{-\half} \right)\ket{0}  \, , 
\end{align}
as well as the state with $J^+ \, \leftrightarrow \, J^-$. Notice that the third term contributes with $T^2$ instead of $T^3$. This is because of the Sugawara relation $L_{-1} \sim J^+_{-\half} J^-_{-\half}$, which allows us to write 
\begin{align}
J^+_{-\half} J^+_{-\half} J^-_{-\half} \ket{0} \sim J^+_{-\half}L_{-1} \ket{0} \in \CV_2 \subset \CV_3 \,,
\end{align}
where $\CV_k$ is defined as in \eqref{eq:su2filter}. Hence, the state given in \eqref{eq:su2null1} lives in $V_2 \equiv \CV_2 / \CV_1$, and contributes with $T^2$ before its removal by the definition of the refined character \eqref{eq:su2refchar}.
\begin{align}
\left( a^2+\frac{1}{a^2} \right)q^{3/2} T^2
\end{align}
from the refined character. At the order $q^2$, we find two null states with $\mathfrak{su}(2)$ weight $J^0 = 0$ and a null state with weight $J^0= \pm 1$ each. Therefore, we remove 
\begin{align}
 q^2 \left(2T^2 + a^2 T^3 + \frac{T^3}{a^2} \right)  \ 
\end{align}
from the refined character upon counting the $T$-weights. We find that the refined character indeed agrees with the lens space index.

Next let us discuss the case with $r=4$ and $m=2$. The lens index after the appropriate limit is given by
\bea
\cI_{(A_1,D_3),\, 4}^{\textrm{Mac}} ( a,2 )
&=& 
1
+\left[\left(1+\frac{1}{a^2}+a^2\right)+\left(2+\frac{1}{a^2}+a^2\right) q^4+\left(2+\frac{1}{a^2}+a^2\right) q^8+\cdots\right] t\nn\\
&{ }&
+\bigg[\left(1+\frac{1}{a^4}+\frac{1}{a^2}+a^2+a^4\right)+\left(2+\frac{1}{a^4}+\frac{2}{a^2}+2 a^2+a^4\right) q^4\nn\\
&{ }&\qquad+\left(5+\frac{2}{a^4}+\frac{4}{a^2}+4 a^2+2 a^4\right) q^8+\cdots \bigg]t^2\nn\\
&{ }&
   +\bigg[\left(1+\frac{1}{a^6}+\frac{1}{a^4}+\frac{1}{a^2}+a^2+a^4+a^6\right)\nn\\
&{ }& \qquad+\left(
   2+\frac{1}{a^6}+\frac{2}{a^4}+\frac{2}{a^2}+2 a^2+2 a^4+a^6\right)
   q^4+\cdots \bigg]t^3 \nn\\
&{ }&+~ \cO\left( t^{4} \right) \,.
\eea
Taking the \emph{round limit}, \ie~rescaling $ q\rightarrow q^{1/4}$ and then taking $t \rightarrow q T$ gives back the round index/usual vacuum character of the chiral algebra. This works with our analysis as $\lambda = \frac{[2m]_r}{r} = 0$ and thus we precisely expect this to happen.

\paragraph{Lens index for $L(5,1)$.}

Next let us discuss a couple of examples for the case with $r=5$. First we take $m=1$ and take the Macdonald limit to find
\begin{align}
\begin{split}
\cI_{(A_1,D_3), \, r=5}^{\textrm{Mac}}( a,1 )
&=
1
+\frac{q^2 t}{a^2}
+a^2 q^3 t
+\frac{q^4 t^2}{a^4}
+2 q^5 t
+q^6 \left(a^4 t^2+\frac{t^3}{a^6}\right)
+\frac{q^7\left(t+t^2\right)}{a^2}\\
&\quad  
+q^8 \left(a^2 t+a^2 t^2+\frac{t^4}{a^8}\right)
+q^9 \left( \frac{t^2}{a^{4}}+ \frac{t^3}{a^{4}}+a^{6}t^3\right)\\
&\quad +\cO\left( q^{10} \right)\,.
\end{split}
\end{align}
Taking the $t \rightarrow 1$ limit, this precisely matches the $\left( \frac{2}{5} ,\frac{3}{5} \right)$-twisted character of $\widehat{\mathfrak{su} ( 2 )}_{-\frac{4}{3}}$, if we take the highest weight/vacuum condition 
\bea
\ket{ h_0 } \ = \ \ket{ - \frac{1}{15} } \,.
\eea
By picking this vacuum we precisely obtain the correct nullvectors. In this example they first appear at order $q^{7/5}$ with contribution
\bea
\frac{q^{7/5} }{a^2} \,.
\eea
We can go to high order in $q$ and compare that the other nullstates appearing with this vacuum precisely make the character match to the correct limit of the lens index.

Let us provide a second example with $r=5$, but now we take $m=2$. After taking the above limit, we find
\begin{align}
\begin{split}
\cI_{(A_1,D_3), \, r=5}^{\textrm{Mac}}( a,2 )
&=
1
+a^2 q t
+a^4 q^2 t^2
+a^6 q^3 t^3
+q^4\left(\frac{t}{a^2}+a^8 t^4\right)
+q^5 \left(2 t+a^{10} t^5\right)\\
&
\quad +q^6 \left(a^2 t+a^2 t^2 + a^{12} t^6 \right)  
+q^7\left(a^4 t^2+a^4 t^3+a^{14} t^7\right)\\
&
\quad +q^8 \left(\frac{t^2}{a^4}+a^6 t^3+a^6 t^4+a^{16} t^8\right)
+q^9\left(\frac{t}{a^2}+\frac{t^2}{a^2}\right)\\
&
\quad + \cO\left( q^{10} \right)\,.
\end{split}
\end{align}
Taking $t \rightarrow 1$ and rescaling $q\rightarrow q^{1/5}$, this precisely matches the $\left( \frac{1}{5} ,\frac{4}{5}\right)$-twisted character of $\widehat{\mathfrak{su}\left( 2 \right)}_{-\frac{4}{3}}$ if we take the vacuum 
\bea
\ket{ h_0 } \ = \ \ket{ - \frac{1}{5} } \,.
\eea
Again we can check that this leads to the correct nullstates to match with the lens index. Also let us mention that at order $q^{9}$ there is the term $a^{18}( q t )^{9}$ missing. This is simply because of the cutoff in our expansion when we calculate the lens index.

\paragraph{Lens index for $L(6,1)$.}

Finally let us write down one case with $r=6$; We shall pick $m=1$. After taking the appropriate limit we get the following result
\bea
\cI_{(A_1,D_3), \, r=6}^{\textrm{Mac}}( a,1 )
&=&
1+\frac{q^2}{a^2}+\left(\frac{1}{a^4}+a^2\right) q^4+\left(\frac{1}{a^6}+2\right) q^6+\left(\frac{1}{a^8}+a^4+\frac{2}{a^2}\right) q^8\nn\\
&&+\left(\frac{1}{a^{10}}+\frac{2}{a^4}+2 a^2\right) q^{10}+\cO\left(q^{11}\right)\,.
\eea
Taking the $t\rightarrow 1$ and $q\rightarrow q^{1/6}$ limits, we find precise agreement with the $\left(  \frac{2}{6}, \frac{4}{6}\right)$-twisted character of $\widehat{\mathfrak{su}\left( 2 \right)}_{-\frac{4}{3}}$, if we take the following vacuum
\bea
\ket{ h_0 } \ = \ \ket{ - \frac{1}{9} } \,.
\eea

\paragraph{General structure.}

In Table~\ref{su2table} we explicitly list the identification of the lens index with the twisted character for cases of ``small" $r$. Among other examples, we have checked those cases as an expansion in $q$ to a very high order and were to identify the correct null vectors to reproduce the lens index.
{\renewcommand{\arraystretch}{1.5}
\smallskip
\begin{table}[h!]
\begin{center}
\begin{tabular}{ c : c | c | c : c}
 $L(p,1)$ & $\left( m_1,m_2 \right)$ & limit & $\left( \lambda_1 , \lambda_2 \right)$-twisted character &  vacuum: $\ket{ h_{0} }$\\
  \hline
 $L(3,1)$ 	& $\left( 1, 2 \right)$ 	& $t\rightarrow 1$ 	& $\left( \frac{2}{3} , \frac{1}{3} \right)$ 	&  $\ket{ \frac{1}{9} }$\\
 		& $\left( 2, 1 \right)$ 	& $t\rightarrow 1$ 	& $\left( \frac{1}{3} , \frac{2}{3} \right)$ 	&  $\ket{ - \frac{1}{9} }$\\
 \hline
 $L(4,1)$	& $\left( 1, 3 \right)$ 	& $t\rightarrow 1$ 	& $\left( \frac{2}{4} , \frac{2}{4}\right)$	&  $\ket{ 0 }$\\
		& $\left( 2, 2 \right)$ 	& round		 	& $\left( 0,0\right)$ 					&  $\ket { 0 }$\\
		& $\left( 3, 1 \right)$ 	& $t\rightarrow 1$ 	& $\left( \frac{2}{4} , \frac{2}{4}\right)$ 	&  $\ket{ 0 }$\\
 \hline
 $L(5,1)$	& $\left( 1, 4 \right)$ 	& $t\rightarrow 1$ 	& $\left( \frac{2}{5} , \frac{3}{5}\right)$	&  $\ket{ -\frac{1}{15} }$\\
		& $\left( 2, 3 \right)$ 	& $t\rightarrow 1$ 	& $\left( \frac{4}{5} , \frac{1}{5}\right)$	&  $\ket{ \frac{1}{5} } $\\
		& $\left( 3, 2 \right)$ 	& $t\rightarrow 1$ 	& $\left( \frac{1}{5} , \frac{4}{5}\right)$	&  $\ket{ -\frac{1}{5} }$\\
		& $\left( 4, 1 \right)$ 	& $t\rightarrow 1$ 	& $\left( \frac{3}{5} , \frac{2}{5}\right)$	&  $\ket{ \frac{1}{15} }$\\
\hline
 $L(6,1)$	& $\left( 1, 5 \right)$ 	& $t\rightarrow 1$ 	& $\left( \frac{2}{6} , \frac{4}{6}\right)$	&  $\ket{ -\frac{1}{9} }$\\
		& $\left( 2, 4 \right)$ 	& $t\rightarrow 1$ 	& $\left( \frac{4}{6} , \frac{2}{6}\right)$	&  $\ket{ \frac{1}{9} }$\\
		& $\left( 3, 3 \right)$ 	& round			& $\left( 0 , 0\right)$					&  $\ket{ 0 }$\\
		& $\left( 4, 2 \right)$ 	& $t\rightarrow 1$	& $\left( \frac{2}{6} , \frac{4}{6}\right)$	&  $\ket{ -\frac{1}{9} }$\\
		& $\left( 5, 1 \right)$ 	& $t\rightarrow 1$	& $\left( \frac{4}{6} , \frac{2}{6}\right)$	&  $\ket{ \frac{1}{9} }$\\
\end{tabular}
\caption{\label{su2table}
We list here examples of the lens index of the $(A_1, D_3)$ Argyres-Douglas theory for a few value of $r$ and their precise identification with the twisted chiral algebra of $\widehat{\mathfrak{su}(2)}_{-\frac{4}{3}}$. The first two columns list the lens index data. We then identify the correct limit in the third column, which allows one to precisely match to a twisted character with $(\lambda, 1-\lambda)$ -- as specified in the fourth column -- and the vacuum/highest-weight prescription given in the rightmost column.}
\end{center}
\end{table}}

Let us make a comment for the general $(A_1, D_{2N+1})$ case, where the chiral algebra is given by $\widehat{\mathfrak{su}(2)}_{-\frac{4N}{2N+1}}$. As we have seen in \eqref{eq:A1D5idx}, the general form of the lens index does not always admit a well-defined ``Schur limit", but rather one has to work with the Macdonald limit. We especially note that there are states of the weight $q^0$. From the chiral algebra perspective, it is possible to obtain these terms by imposing $J^0_0 \ket{\textrm{vac}} \neq 0$. As we will see shortly in Section~\ref{sec:su3}, the seemingly unusual condition on the vacuum (more properly highest-weight state) gives rise to a highly non-trivial match between the (Macdonald limit of the) lens index and the (refined) character of the corresponding module. It would be interesting to find such a condition for general $N>1$, akin to the $N=1$ case. We leave this for future work. 

\subsection{The $\left( A_1,D_4 \right)$ theory and its twisted chiral algebra} \label{sec:su3}

In this Section, we will focus on the $(A_1, D_4)$ Argyres-Douglas theory, for which the protected chiral algebra is given by the affine Kac-Moody algebra $\widehat{\mathfrak{su}(3)}_{-\frac{3}{2}}$. We shall now extend this to the case with non-trivial twisting.\footnote{As it is clear which flavor symmetry we are talking about here, we shall remove the subscript for the discrete holonomies, and simply write $m \equiv m_{a}$ in the following.}

\subsubsection{The twisted $\widehat{\mathfrak{su}(3)}_{-\frac{3}{2}}$ chiral algebra}

Let us now introduce the twisted $\widehat{\mathfrak{su}(3)}_{-\frac{3}{2}}$ chiral algebra. We proceed as before by first introducing a twisting action that is consistent with the OPEs of the (untwisted) chiral algebra and then compute the (refined) character for the twisted module. 

We shall first introduce the standard generators for the $\widehat{\mathfrak{su}(3)}_k$ Kac-Moody algebra normalized such that 
\bea
 \cK\left( J^{a}, J^{b} \right) & = & \tr_{\mathrm{fund}} \left( J^{a} J^{b} \right) \ = \ \delta^{ab} \,.
\eea
The OPEs corresponding to the standard standard basis of the $\widehat{\mathfrak{su}(3)}_{k}$ algebra read
\bea
J^{a}\left( z \right)J^{b}\left( w \right) & = & \frac{k \delta^{ab}}{\left( z-w \right)^{2}}+\frac{f^{ab}{}_{c} J^{c}\left( w \right)}{\left( z-w \right)}\,,
\eea
where $a,b \in\{1, \ldots , 8\}$ are the generators and $f^{abc}$ are the usual structure constants of $SU(3)$. We shall put this into a more convenient form by defining 
\beaa
H^{1} & \ = \  \frac{1}{2} \left(  \frac{J^{8}}{\sqrt{3}} + J^{3} \right) \,, \quad  
& H^{2} & \ = \ \frac{1}{2} \left(  \frac{J^{8}}{\sqrt{3}} - J^{3} \right)\,, \quad  
& H^{3} & \ = \ - \frac{J^{8}}{\sqrt{3}}\,, \\
E^{1,\pm}  & \ =  \  \frac{1}{2}\left( J^{6} \pm \ii J^{7} \right) \,, \quad
& E^{2,\pm} & \ = \  \frac{1}{2}\left( J^{4} \mp \ii J^{5} \right) \,, \quad
& E^{3,\pm} & \ = \ \frac{1}{2}\left( J^{1} \pm \ii J^{2} \right) \,.
\eeaa
Of course by doing so, we have introduced a spurious Cartan generator and we shall account for that in the following. The OPEs then read
\bea\label{su3ope1}
 H^{i}\left( z \right)E^{j,\pm}\left( w \right) & \ \sim  \ & \pm m^{ij} \frac{E^{j,\pm}\left( w \right)}{z-w} \,, \quad m_{ij} \ = \ \left( \begin{array}{c c c} 0 & 1 & -1 \\ -1 & 0 & 1 \\ 1 & - 1 & 0 \end{array} \right) \,,
\eea
as well as
\bea\label{su3ope2}
H^{i}\left( z \right)H^{j}\left( w \right) & \sim & \frac{\widehat A^{ij} k}{3 (z-w)^{2}}\,,
\eea
where $\widehat A^{ij}$ is the (affine) Cartan matrix of $\widehat{\mathfrak{su}(3)}_{k}$
and finally
\bea
E^{j,+}\left( z \right)E^{j,-}\left( w \right) & \sim & \frac{k}{(z-w)^{2}} + \frac{\alpha^{j, \pm} \cdot H}{z-w}\,, \label{su3ope3} \\
E^{i,\pm}\left( z \right)E^{j,\pm}\left( w \right) & \sim &\pm \epsilon^{ijk} \frac{E^{k,\mp}}{z-w} \,,\label{su3ope4}
\eea
where we have denoted by $\alpha_{k}^{j, \pm}\cdot H = \alpha_{k}\left( E^{j, \pm} \right)H^{k}$, where $\alpha_k\left( E^{j,\pm} \right)$ denotes the k$^{\rm th}$ root of the corresponding generators.

Now let us proceed to introduce the twisting action relevant for comparing with the lens index. From the OPEs it is clear that we are only allowed the following twist
\beaa
  H^{j} \left( z \right) & \ \rightarrow   \ && H^{j} \left( e^{2\pi \ii} z \right) \ = \  H^{j}\left( z \right) \,,\\
  E^{j, \pm} \left( z \right) & \  \rightarrow  \  &&  E^{j, \pm} \left( e^{2\pi \ii} z \right) \ = \ e^{\pm 2\pi \ii \lambda_j} E^{j,\pm}\left( z \right)\,,
\eeaa
where $\lambda_3 = - \lambda_1 -\lambda_2 \mod \mathbb{Z}$. As usual we shall choose $\lambda_i \in [0,1)$ and therefore $\lambda_3$ is uniquely fixed. Thus our twisted index is defined by \emph{two} constants $\lambda_{1,2}  \in [0,1) $.

As before we can now introduce mode expansions for the twisted fields
\bea
E^{j, \pm} \left( z \right) & = & \sum_{m \in \mathbb{Z}} E^{j, \pm}_{m \mp \lambda_j} z^{-m -1 \pm \lambda_j} \,,
\eea
and we can compute the commutation relations of the modes from the OPEs above to obtain
\bea
\left[H^{i}_{m},H^{j}_{n}\right] & \ = \ & \frac{m k}{3} \hat{A}^{ij} \delta_{m+n,0} \,,\\
\left[H^{i}_{m}, E^{j,\pm}_{n \mp \lambda_j}\right] & \ = \ & m_{ij} E^{j, \pm}_{m+n \mp \lambda_j} \,,\\
\left[E^{i, \pm}_{m \mp \lambda_i}, E^{j,\pm}_{n \mp \lambda_j}\right] & \ = \ & \pm\epsilon^{ij}{}_{k} E^{k, \mp}_{m+n \mp \lambda_k}\,,\\
\left[E^{i, \pm}_{m\mp \lambda_i}, E^{i,\mp}_{n \pm \lambda_i}\right] & \ = \ & k \delta_{m+n} \left( m \mp \lambda_i \right) +\alpha_{k}^{i, \pm}\cdot H \,,
\eea
where we have adopted the notation used for the OPEs in~\eqref{su3ope1}-\eqref{su3ope4}. With this we can read off the twisted characters by counting states and taking into account possible null states along the way. 

Now depending on the chosen twisting, we define the (twisted) vacuum of the $\widehat{\mathfrak{su}( 3 )}_{k}$ chiral algebra by the following data
\bea\label{vacuasu3}
\big|\mathrm{vac}\big\rangle & = & \ket{ \left( h_{0}^{1},h_{0}^{2},h_{0}^{3} \right); e^{\pm, k}_{0} } : \ \left\{ 
\begin{array}{ll}
E_{\nu}^{i, \pm} |\mathrm{vac}\rangle \ = \ 0  \,, \ & \nu > 0, \  i = 1,2,3\,,\\
H^{i}_{\nu} |\mathrm{vac}\rangle \ = \ 0  \,, \ & \nu > 0, \  i = 1,2,3\,,\\
H^{i}_{0} |\mathrm{vac}\rangle \ = \ h^{i}_{0}|\mathrm{vac}\rangle  \,, \ & i = 1,2,3\,, \\
E_{0}^{i_0, \pm} |\mathrm{vac}\rangle \ = \ e_{0}^{i_0, \pm} |\mathrm{vac}\rangle  \,, \ & \text{iff} \ \lambda_{i_0} =0,\ i_0 \in \{1,2,3\}\,,\\
\end{array}
\right.
\eea
where $h^{i}_{0}, e_{0}^{i_0, \pm} \in \mathbb{R}$. Note especially the fourth condition we impose on the ``vacuum"; this seemingly unusual condition is imposed if and only if one (and only one) of the twist parameters $\lambda_i$ become zero. We shall discuss it further in some detail below.

\subsubsection{Comparison with the lens index}

We shall now compare the twisted character of $\widehat{\mathfrak{su}( 3 )}_{-\frac{3}{2}}$ against the lens index of the $\left( A_1, D_4 \right)$ Argyres-Douglas theory computed in Section~\ref{secA1D4}. Let us first state the general result and then go over a few examples. 

We start by removing the following prefactor from the lens index
\bea\label{Eqn:I0SU3}
\cI^{(0)}_{r, \, (\lambda_1, \lambda_2, \lambda_3)} = \left( \frac{pq}{t} \right)^{\frac{\cA\left( r; \lambda_1, \lambda_2,\lambda_3 \right)}{2}} \,,
\eea
where
\bea
\cA\left( r; \lambda_1, \lambda_2,\lambda_3 \right) & = & \frac{ [r \lambda_1 \cdot r \lambda_2]_r}{r}+\frac{[r \lambda_1 \cdot r \lambda_3]_r}{r}+\frac{[r \lambda_2 \cdot r \lambda_3]_r }{r} \,.
\eea
The Macdonald index is then obtained as before via\footnote{Recall that $\CI^{(0)}(p, q, t)$ is the zero-point energy of the $(A_1,D_4)$ Argyres-Douglas theory. In particular it is precisely given by $\cI^{(0)}_{r, \, \lambda}$ in equation~\eqref{Eqn:I0SU3} upon identification of $(\lambda_1,\lambda_2)$ as given in~\eqref{Eqn:lambdavsmSU3}.}
\bea
\CI^{\textrm{Mac}}(q, t) = \lim_{p \to 0} \left[\CI(p, q, t)/\CI^{(0)}(p, q, t)\right]  \,.
\eea
We obtain the general relation
\begin{align}
 \chi^{\textrm{ref}}_{\CV_{\ket{\textrm{vac}}}^{(\lambda_1, \lambda_2)}} (a; q, T) = \begin{dcases}
  \CI^{\textrm{Mac}}_{(A_1, D_4), r} (a, m_i \neq 0; q\to q^{1/r}, t \to T) \, , & (\lambda_1, \lambda_2) \neq (0, 0)\,, \\
  \CI^{\textrm{Mac}}_{(A_1, D_4), r} (a, m_i = 0; q\to q^{1/r}, t \to qT) \, , & (\lambda_1, \lambda_2) = (0, 0) \,.
 \end{dcases}
\end{align}
Notice that the identification of the lens index with the character depends on the choice of the twisting parameters $\lambda_i$ or the discrete holonomies $m_i$ as in the other cases. 

The precise identification of the lens index with the twisted character involves a few more identifications of parameters. Firstly, matching the discrete holonomies $(m_1,m_2)$ of the lens index to the twisting parameters is slightly nontrivial, as we redefine the natural fugacities in the computation of the lens index of $\left( A_1,D_4 \right)$ Argyres-Douglas theories (see equations~\eqref{Eqn:redefA1D4a} and~\eqref{Eqn:redefA1D4m}). Studying various examples, we find that
\begin{align}\label{Eqn:lambdavsmSU3}
\lambda_1 \ = \ \frac{\left[m_1+2m_2\right]_{r}}{r} \,,\qquad 
\lambda_2 \ = \ \frac{\left[-2 m_1-m_2\right]_{r}}{r} \,,\qquad
\lambda_3 \ = \ \frac{\left[m_1-m_2\right]_{r}}{r} \,,
\end{align}
gives the correct identification of the twisting parameters in terms of the discrete holonomies of the $\left( A_1,D_4 \right)$ Argyres-Douglas lens index.

Secondly, we need to impose conditions on the ``vacuum" state (or more precisely the highest-weight state). 
The vacuum conditions appearing in~\eqref{vacuasu3} are shifted to
\bea
h_{0}^{1}  \ = \ \frac{1}{2} \left( \lambda_3 - \lambda_2\right) \,, \qquad h_{0}^{2}  \ = \ \frac{1}{2} \left( \lambda_1 - \lambda_3\right) \,, \qquad h_{0}^{3}  \ = \ \frac{1}{2} \left( \lambda_2 - \lambda_1\right) \,,
\eea
and, if one of the twisting parameters vanishes, we have a further ``vacuum" condition given by
\bea
e^{j,+}_{0}e^{j,-}_{0} = -\tfrac{1}{2} \,, \quad \text{if} \quad \lambda_{j} \equiv 0 \ \text{ and } \ \lambda_i \neq 0 \,, \ \forall i \neq j \,.
\eea
We have checked these rules in a large amount of examples, and find precise agreement to rather high order in $q$- and $(q,t)$-expansions. 

Let us make some remarks on the ordinary character limit $T \to 1$ of the refined character. There are the same two limits as in the $\widehat{\mathfrak{su}( 2 )}_{k}$ case, but there is also a limit, which we dubbed \emph{mixed}, where taking $t\rightarrow 1$ \emph{does not} in fact lead to a twisted character. In this case one is forced to use the refined character, since there is no proper unrefined limit we can take. This mixed limit appears as soon as one and only one of the twisting parameter $\lambda_i$ vanishes. Together we find the following identification
\bea
\chi_{\CV^{(\lambda_1, \lambda_2)}_{\ket{\mathrm{vac}}}} (q) \ = \ 
\begin{dcases}
 \lim_{t\rightarrow 1} \CI^{\textrm{Mac}}_{(A_1, D_4), r}(a, m \neq 0; q \to q^{1/r}, t) \,, & \lambda_{i} \neq 0 \,, \ \forall i = 1,2,3\,,\\ 
\lim_{t\rightarrow q} \CI^{\textrm{Mac}}_{(A_1, D_4), r}(a, m=0; q \to q^{1/r}, t) \,, & \lambda_{i} = 0 \,, \ \forall i = 1,2,3\,,
\end{dcases}
\eea
with the \emph{ordinary} characters. If there is exactly one $i_0 \in \{1,2,3\}$ with $\lambda_{i_0} = 0$, we do not get any identification with the ``unrefined" character but we are still able to precisely match with the \emph{refined} character.\footnote{Notice that when two of the $\lambda_i$'s are zero, it becomes untwisted since $\sum_{i=1}^{3} \lambda_i = 0$.} 


\subsubsection{Examples of the lens index for the $\left( A_1, D_4 \right)$ theory}\label{A1D4examples}

Let us discuss some examples of the lens index of the $\left( A_1, D_4 \right)$ Argyres-Douglas theory. We shall elaborate in some detail how to identify the lens index (in the limits discussed above) with the appropriately twisted characters of $\widehat{\mathfrak{su}( 3 )}_{-\frac{3}{2}}$.\footnote{As it is clear which flavor symmetry we are talking about here, we shall remove the subscript for the discrete holonomies, and simply write $m_{i} \equiv m_{a}^{(i)}$ in the following.}

\paragraph{Lens index for $L(4,1)$ with $(m_1, m_2)=(1, 3)$.}

Let us start with an example for $r=4$. We pick $(m_1,m_2)=(1,3)$, and find after taking the appropriate limit
\begin{align}
\begin{split}
\cI_{( A_1,D_4 ), \, 4}^{\textrm{Mac}} &\left( ( a_1 ,1 ),( a_2 ,3 ) \right)  \\
 &= 1
   +q t \left[\frac{1}{{a_1}^2 {a_2}}+{a_1} {a_2}^2\right] 
   +q^2
   \left[\left(\frac{{a_1}}{{a_2}}+\frac{{a_2}}{{a_1}}\right) t+\left(\frac{1}{{a_1}^4 {a_2}^2}+{a_1}^2 {a_2}^4\right) t^2\right]\\
 & \quad +q^3 \left[\left(\frac{1}{{a_1} {a_2}^2}+{a_1}^2 {a_2}\right) t+\left(\frac{1}{{a_1}^3}+{a_2}^3\right)
   t^2+\left(\frac{1}{{a_1}^6 {a_2}^3}+{a_1}^3 {a_2}^6\right) t^3\right]
   \\
 & \quad +q^4 \bigg[3 t+\left(\frac{1}{{a_1}^3
   {a_2}^3}+\frac{{a_1}^2}{{a_2}^2}+\frac{{a_2}^2}{{a_1}^2}+{a_1}^3 {a_2}^3\right) t^2+\left(\frac{1}{{a_1}^5 {a_2}}+{a_1}
   {a_2}^5\right) t^3\\
 &  \quad   \qquad+\left(\frac{1}{{a_1}^8 {a_2}^4}+{a_1}^4 {a_2}^8\right) t^4\bigg]
      \\
  & \quad +q^5
   \bigg[\left(\frac{1}{{a_1}^2 {a_2}}+{a_1} {a_2}^2\right) t+\left({a_1}^3+\frac{1}{{a_2}^3}+\frac{2}{{a_1}^2 {a_2}}+2 {a_1}
   {a_2}^2\right) t^2\\
 & \quad \qquad+\left(\frac{1}{{a_1}^5 {a_2}^4}+\frac{{a_2}}{{a_1}^4}+\frac{{a_2}^4}{{a_1}}+{a_1}^4 {a_2}^5\right)
   t^3\bigg]\\
 & \quad +\cO\left( q^{6} \right)\,.
\end{split}
\end{align}
Let us first compute the refined character for the $(\lambda_1, \lambda_2)$-twisted Verma module. As in the case of \eqref{eq:su2verma}, this can easily be computed by enumerating generators and relations inside the plethystic exponential. This gives
\begin{align}
 \chi^{\textrm{ref}}_{\CV_{\textrm{Verma}}^{(\lambda_1, \lambda_2)}}
  = \textrm{PE} \left[\frac{T}{1-q} \left(2q + \sum_{i=1}^3 \left( q^{\lambda_i} z^{{\a}_i} +q^{1-\lambda_i} z^{-{\a}_i} \right) \right) + \frac{qT}{1-q} - \frac{q T^2}{1-q}  \right]_{q, T, {z}} \, , 
\end{align}
where we have used a short-hand notation ${z}^{{\a}} \equiv \prod_j z_j^{\a_j}$. The first term inside the parenthesis comes from the $\widehat{\mathfrak{su}(3)}$ generators and the last two comes from the Virasoro generator and the Sugawara relation. The refined character for the $(\frac{3}{4}, \frac{3}{4})$-twisted Verma module reads
\begin{align}
\begin{split}
 \chi^{\textrm{ref}}_{\CV_{\textrm{Verma}}^{(\frac{3}{4}, \frac{3}{4})}}
  &= 
  1 + q^{\frac{1}{4}} T\left[\frac{1}{a_1^2 a_2} + a_1 a_2^2\right] 
  +q^{\half} \left[\left(\frac{a_1}{a_2}+\frac{a_2}{a_1}\right) T + \left(a_1^2 a_2^4+\frac{a_2}{a_1}+\frac{1}{a_1^4 a_2^2}\right) T^2 \right] \\
&{ }\qquad + q^{\frac{3}{4}} \Bigg[ \left(\frac{1}{a_1 a_2^2} + a_1^2 a_2 \right) T +\left(a_2^3+a_1^2 a_2+\frac{1}{a_1^3}+\frac{1}{a_1 a_2^2}\right) T^2\\
&{ }\qquad \qquad \qquad + \left(a_1^3 a_2^6+a_2^3+\frac{1}{a_1^3}+\frac{1}{a_1^6 a_2^3}\right) T^3 \Bigg] \\
&{ } \qquad  + \cO(q^1) \, . 
\end{split}
\end{align}
We see that the lens space index agrees with the twisted character for the Verma module upon $q \to q^{1/4}$ and $t \to T$ up to some extra terms. These extra terms are removed by choosing the vacuum state to be
\bea\label{Eqn:vacSU3r4m13}
 \ket{ \left( h_{0}^{1},h_{0}^{2},h_{0}^{3} \right); e^{\pm, k}_{0}}  & = & \ket{ \left( - \frac{1}{8},\frac{1}{8}, 0 \right) ;  \varnothing } \,,
\eea
where $h_0^{3}=-h_0^{1}-h_0^{2}$. For example, we obtain a null state at level $\half$ given as
\begin{align}
 \left( E_{-\frac{1}{4}}^{1, +} \, E_{-\frac{1}{4}}^{2, +} - \frac{1}{2} E_{-\frac{1}{2}}^{3, -} \right) \ket{\textrm{vac}} \, ,
\end{align}
which removes the term 
\bea
\frac{q^{\half} T^2 a_2}{a_1}
\eea
 from the character.\footnote{In our convention, the root vectors are chosen so that the charges for the generators are given as 
\bea
E^{1, +} \to \frac{1}{a_1^2 a_2}\,,\quad E^{2, +} \to a_1 a_2^2\,,\quad E^{3, +} \to \frac{a_1}{a_2}\,.
\eea} We get four null states at level $\frac{3}{4}$ given by 
\begin{align}
\begin{split}
 \left( -\frac{3}{2} E^{2, +}_{-\frac{1}{4}} E^{3, -}_{-\half} + E^{1, +}_{-\frac{1}{4}} E^{2, +}_{-\frac{1}{4}} E^{2, +}_{-\frac{1}{4}} \right) \ket{\textrm{vac}} \, , 
& \quad \left( -\frac{1}{2} E^{1, +}_{-\frac{1}{4}} E^{3, -}_{-\half} + E^{1, +}_{-\frac{1}{4}} E^{1, +}_{-\frac{1}{4}} E^{2, +}_{-\frac{1}{4}} \right) \ket{\textrm{vac}} \\
 \left( E^{2, +}_{-\frac{1}{4}} E^{3, +}_{-\half} -\half E^{1, -}_{-\frac{3}{4}} \right) \ket{\textrm{vac}} \, , 
& \quad \left( E^{1, +}_{-\frac{1}{4}} E^{3, +}_{-\half} +\half E^{2, -}_{-\frac{3}{4}}\right) \ket{\textrm{vac}} \,.
\end{split}
\end{align}
These null states remove 
\bea
q^{\frac{3}{4}}T^3 a_2^3 \,,\qquad \frac{q^{\frac{3}{4}} T^3}{a_1^3} \,,\qquad q^{\frac{3}{4}} T^2 a_1^2 a_2 \,,\quad \text{and}\quad \frac{q^{\frac{3}{4}} T^2}{a_1 a_2^2} \,,
\eea
 respectively from the Verma module character. In the $T \to 1$ limit, we recover the $\left( \frac{3}{4}, \frac{3}{4} \right)$-twisted (ordinary) character of $\widehat{\mathfrak{su}( 3 )}_{-\frac{3}{2}}$ with highest weight/vacuum prescription given in~\eqref{Eqn:vacSU3r4m13}.

\paragraph{Lens index for $L(4,1)$ with $(m_1, m_2)=(1, 1)$.}

As a second example let us discuss a case where one has to look at the~\emph{mixed} limit of the lens index, in order to match it to a refined character. We still keep $r=4$, and pick $\left( m_1,m_2 \right) =\left( 1,1 \right) $. The corresponding lens index in the Macdonald limit is then given by
\begin{align}
\begin{split}
\cI_{( A_1,D_4 ), \, 4}^{\textrm{Mac}}& \left( ( a_1 ,1 ),( a_2 ,1 ) \right)\\
& = 1
+\left(\frac{{a_1}}{{a_2}}+\frac{{a_2}}{{a_1}}+1\right)t
+\left(\frac{{a_1}^2}{{a_2}^2}+\frac{{a_2}^2}{{a_1}^2}+\frac{{a_1}}{{a_2}}+\frac{{a_2}}{{a_1}}+1\right)t^2+ \cdots  \\
& \quad
+ \left[ \left({a_1}^2 {a_2}+{a_1} {a_2}^2\right)t+\left({a_1}^3+{a_1}^2 {a_2}+{a_1} {a_2}^2+{a_2}^3\right) t^2+ \cdots  \right] q \\
& \quad
+ \left[ \left({a_1}^4 {a_2}^2+{a_1}^3 {a_2}^3+{a_1}^2 {a_2}^4\right)t^2 + \cdots \right]q^2\\
& \quad
+
\left[
\left(\frac{1}{{a_1}^2 {a_2}}+\frac{1}{{a_1} {a_2}^2}\right)t
+\left(\frac{1}{{a_1}^3}+\frac{1}{{a_1}^2 {a_2}}+\frac{1}{{a_1} {a_2}^2}+\frac{1}{{a_2}^3}\right)t^2
+ \cdots
\right]q^3\\
& \quad
+\left[ 
\left(\frac{{a_1}}{{a_2}}+\frac{{a_2}}{{a_1}}+3\right)t
+\left(\frac{{a_1}^2}{{a_2}^2}+\frac{{a_2}^2}{{a_1}^2}+\frac{3 {a_1}}{{a_2}}+\frac{3 {a_2}}{{a_1}}+3\right)t^2
+ \cdots
\right]q^4\\
& \quad
+\left[  \left({a_1}^2 {a_2}+{a_1} {a_2}^2\right)t
+\left(2 {a_1}^3+4 {a_1}^2 {a_2}+4 {a_1} {a_2}^2+2 {a_2}^3\right) t^2 
+ \cdots
\right]q^5 \\
& \quad +\cO\left( q^{6} \right) \,.
\end{split}
\end{align}
Since the $q^{0}$-order goes as an infinite series in $t$, the $t\rightarrow 1$ limit is not well-defined. In particular the states that appear at this order are ``supposed" to appear only at order $q^{4}$ (or rather $q^{1}$ after rescaling $q\rightarrow q^{1/4}$). So the structure of these states is reminiscent of the lens indices where the round limit is the appropriate limit one should take to compare to the chiral algebra. Therefore we have called this case \emph{mixed}. As none of the strict limits give a character, we have to resort to the {refined} version of the character. 

Either way, after rescaling $q\rightarrow q^{1/4}$, we can carefully analyse the spectrum. We find that the above index precisely reproduces the refined $\left( \frac{3}{4}, \frac{1}{4}, 0 \right)$-twisted character of $\widehat{\mathfrak{su}( 3 )}_{-\frac{3}{2}}$, with vacuum given by
\bea\label{Eqn:vacSU3r4m11}
 \ket{ \left( h_{0}^{1},h_{0}^{2},h_{0}^{3} \right); e^{\pm, k}_{0} } & = & \ket{ \left( - \frac{1}{8},\frac{3}{8}, -\frac{1}{4} \right) ;  e_{0}^{+,3} e_{0}^{-,3} = -\frac{1}{2} } \,.
\eea
The first nullstates we can observe in this case have contributions at order $q^{2/4}t$ (they are part of a null space which give also contributions at order $q^{5/4}t^{2}$). In particular the first contribution is precisely given by
\bea
\left(a_2 a_1^2+a_2^2 a_1\right) q^{1/4} t^2 \,.
\eea
At higher order in $q$, but second order in $t$ we further find the following contributions of null states
\bea 
\begin{split}
&\left(\frac{1}{a_1 a_2^2}+\frac{1}{a_1^2 a_2}\right) q^{3/4} t^2\,,&\\
&\left(2 \frac{ a_2}{a_1}+2\frac{ a_1}{a_2}+3\right)q t^2 \,,&\\
&2 \left(a_2 a_1^2+a_2^2 a_1\right) q^{5/4} t^2\,,& \ \mathrm{etc.}\,.
\end{split}
\eea
We can explicitly check that these null vectors are precisely necessary to match the $\left( \frac{3}{4}, \frac{1}{4}, 0 \right)$-twisted character with highest weight/vacuum in equation~\eqref{Eqn:vacSU3r4m11} with the lens index.

\paragraph{Lens index for $L(5,1)$.}

Let us present one case with $r=5$. We shall pick $\left( m_1,m_2 \right) = \left( 3,3 \right)$. After taking the  Macdonald limit, we find the following lens index
\begin{align}
\begin{split}
\cI_{( A_1,D_4 ), \, 5}^{\textrm{Mac}}& \left( ( a_1 ,3 ),( a_2 ,3 ) \right)\\
&=
1+ 
\left(\frac{{a_1}}{{a_2}}+\frac{{a_2}}{{a_1}}+1\right)t
+\left(\frac{{a_1}^2}{{a_2}^2}+\frac{{a_2}^2}{{a_1}^2}+\frac{{a_1}}{{a_2}}+\frac{{a_2}}{{a_1}}+1\right)t^2 \\ 
&\quad
+\left[
 \left({a_1}^2 {a_2}+{a_1} {a_2}^2\right)t
 + \left({a_1}^3+{a_1}^2 {a_2}+{a_1} {a_2}^2+{a_2}^3\right)t^2+\cdots
 \right]q  \\ 
 &\quad + \left[\left({a_1}^4 {a_2}^2+{a_1}^3 {a_2}^3+{a_1}^2 {a_2}^4\right)t^2 +\cdots\right]q^2\\
&\quad + \left[
 \left(\frac{1}{{a_1}^2 {a_2}}+\frac{1}{{a_1} {a_2}^2}\right)t
 + \left(\frac{1}{{a_1}^3}+\frac{1}{{a_1}^2 {a_2}}+\frac{1}{{a_1} {a_2}^2}+\frac{1}{{a_2}^3}\right)t^2 +\cdots\right]q^4\\ 
&\quad + \left[
t \left(\frac{{a_1}}{{a_2}}+\frac{{a_2}}{{a_1}}+3\right)
+t^2 \left(\frac{{a_1}^2}{{a_2}^2}+\frac{{a_2}^2}{{a_1}^2}+\frac{3 {a_1}}{{a_2}}+\frac{3 {a_2}}{{a_1}}+3\right)+\cdots\right]q^5\\
&\quad+ \cO\left( q^{6} \right)\,.
\end{split}
\end{align}
As in the previous example we have to look at the refined character in order to match it with the lens index. Again, after rescaling $q\rightarrow q^{1/5}$, we find that the above index precisely reproduces the refined $\left( \frac{4}{5}, \frac{1}{5}, 0 \right)$-twisted character of $\widehat{\mathfrak{su}( 3 )}_{-\frac{3}{2}}$, with the vacuum 
\bea
 \ket{ \left( h_{0}^{1},h_{0}^{2},h_{0}^{3} \right); e^{\pm, k}_{0} } & = & \ket{ \left( - \frac{1}{10},\frac{2}{5}, -\frac{3}{10} \right) ;  e_{0}^{+,3} e_{0}^{-,3} = -\frac{1}{2} } \,.
\eea
In this case we find the following low order null state contributions
\bea
\begin{split}
&\left(a_2 a_1^2+a_2^2 a_1\right) q^{1/5} t^2\,,&\\
&\left(\frac{1}{a_1 a_2^2}+\frac{1}{a_1^2 a_2}\right) q^{4/5} t^2\,, &\ \mathrm{etc.}\,.
\end{split}
\eea

\paragraph{Lens index for $L(6,1)$.}

Let us close this Section by detailing a case with $r=6$, and $\left( m_1,m_2 \right)=\left( 2,0 \right)$. After taking the appropriate limit, the lens index reads
\begin{align}
\begin{split}
\cI_{( A_1,D_4 ), \, 6}^{\textrm{Mac}}&\left( ( a_1 ,2 ),( a_2 ,0 ) \right)\\
&=
1+\left[\frac{1}{a_1 a_2^2}+\frac{a_2}{a_1}+a_1^2 a_2\right] q^2 t\\
&\quad+\left[\left(\frac{1}{a_1^2 a_2}+\frac{a_1}{a_2}+a_1a_2^2\right) t
+\left(\frac{1}{a_1^2 a_2^4}+\frac{a_2^2}{a_1^2}+a_1^4 a_2^2\right) t^2\right]q^4 \\
&\quad
+\left[3t+\left(\frac{1}{a_1^3}+a_1^3+\frac{1}{a_2^3}+\frac{1}{a_1^3 a_2^3}+a_2^3+a_1^3 a_2^3\right) t^2\right]q^6 
+\cO\left( q^{8} \right)\,.
\end{split}
\end{align}
After taking the limit $q\rightarrow q^{1/6}$, and $t\rightarrow T$, this precisely matches with the $\left( \frac{2}{6}, \frac{2}{6},\frac{2}{6} \right)$-twisted refined character of $\widehat{\mathfrak{su}( 3 )}_{-\frac{3}{2}}$, given the following vacuum
\bea
 \ket{ \left( h_{0}^{1},h_{0}^{2},h_{0}^{3} \right); e^{\pm, k}_{0} } & = & \ket{ \left(  0,0,0 \right) ;  \varnothing } \, .
\eea
It reproduces the ordinary character in the limit $T \to 1$ as expected.

\paragraph{Matching for more general cases.}

Finally in Table~\ref{tablesu3} we write down the explicit matching of a large class of examples of $\left( \lambda_1,\lambda_2 \right)$-twisted characters of $\widehat{\mathfrak{su}( 3 )}_{-\frac{3}{2}}$, with the appropriate limits of the $\left( A_1,D_4 \right)$ Argyres-Douglas lens indices, $\cI_{( A_1,D_4 ), \, r}^{\textrm{Mac}}( a_j ,m_j )$. 

\begin{table}[h!]
\begin{center}
{\renewcommand{\arraystretch}{1.13}
\noindent
\begin{tabular}{ c : c | c | c : c}
$L(p,1)$ & $\left( m_1, m_2 \right)$ & Limit & $\left( \lambda_1 , \lambda_2 , \lambda_3 \right)$ &  Vacuum: $\big| \left( h_{0}^{1},h_{0}^{2},h_{0}^{3} \right); e^{\pm, k}_{0}\rangle$\\
  \hline
L(3,1) 
& $\left( 2 , 2 \right),\left( 3 , 3 \right)$ & round 		& $\left( 0 , 0 , 0 \right)$ 	& $\big| \left( 0 , 0 ,0 \right); \varnothing \rangle$\\
\hdashline
& $\left( 1 , 3 \right),\left( 0 , 2 \right),\left( 4 , 0 \right)$ 	& $t \rightarrow 1$ & $\left( \tfrac{1}{3} , \tfrac{1}{3} , \tfrac{1}{3} \right)$ 	& $\big| (0,0,0); \varnothing \rangle$\\
\hdashline
& $\left( 0 , 4 \right),\left( 2 , 0 \right),\left( 3 , 1 \right)
$ 	& $t \rightarrow 1$ & $\left( \tfrac{2}{3} , \tfrac{2}{3} , \tfrac{2}{3}\right)$ 	& $\big| \left( 0 , 0 ,0 \right); \varnothing \rangle$\\
\hline
L(4,1) 	
& $\left( 3 , 1 \right)$ 	& $t \rightarrow 1$  & $\left( \tfrac{1}{4} , \tfrac{1}{4} , \tfrac{2}{4} \right)$ & $\big| \left(  \tfrac{1}{8},-\tfrac{1}{8},0 \right); \varnothing \rangle$\\
& $\left( 1 , 3 \right)$ 	& $t \rightarrow 1$  & $\left( \tfrac{3}{4} , \tfrac{3}{4}, \tfrac{2}{4} \right)$ & $\big| \left(  -\tfrac{1}{8},\tfrac{1}{8},0 \right); \varnothing \rangle$\\
\hdashline
& $\left( 1 , 1 \right)$ 	& mixed  & $\left( \tfrac{3}{4},\tfrac{1}{4},0 \right)$ & $\big| \left(-\tfrac{1}{8},\tfrac{3}{8}, - \tfrac{1}{4} \right); e_{0}^{+,3}e_{0}^{-,3}= -\tfrac{1}{2}  \rangle$\\
& $\left( 3 , 3 \right)$ 	& mixed  & $\left( \tfrac{1}{4},\tfrac{3}{4},0 \right)$ & $\big| \left(-\tfrac{3}{8},\tfrac{1}{8}, \tfrac{1}{4}\right); e_{0}^{+,3}e_{0}^{-,3}=-\tfrac{1}{2}  \rangle$\\
\hdashline
& $\left( 2 , 0 \right),\left( 2 , 4 \right)$ 	& mixed  & $\left( \tfrac{2}{4},0,\tfrac{2}{4} \right)$ & $\big| \left(\tfrac{1}{4},0, \tfrac{1}{4} \right); e_{0}^{+,2}e_{0}^{-,2}=-\tfrac{1}{2}  \rangle$\\
& $\left( 2 , 2 \right)$ 	& mixed  &  $\left( \tfrac{2}{4},\tfrac{2}{4},0 \right)$ & $\big| \left(-\tfrac{1}{4},\tfrac{1}{4},0  \right); e_{0}^{+,1}e_{0}^{-,1}=-\tfrac{1}{2}  \rangle$\\
& $\left( 4 , 2 \right)$ 	& mixed  & $\left( 0,\tfrac{2}{4},\tfrac{2}{4} \right)$ & $\big| \left(0,-\tfrac{1}{4}, \tfrac{1}{4} \right); e_{0}^{+,2}e_{0}^{-,2}=-\tfrac{1}{2}  \rangle$\\
\hline
L(5,1) 	
& $\left(2,0 \right)$ & $t \rightarrow 1$  & $\left(\tfrac{2}{5},\tfrac{1}{5},\tfrac{2}{5}\right)$ & $\big|\left(\tfrac{1}{10},0,-\tfrac{1}{10} \right); \varnothing\rangle$\\
& $\left(4,0 \right)$ & $t \rightarrow 1$  & $\left(\tfrac{4}{5},\tfrac{2}{5},\tfrac{4}{5}\right)$ & $\big|\left(\tfrac{1}{5},0,-\tfrac{1}{5} \right); \varnothing\rangle$\\
& $\left(6,0 \right)$ & $t \rightarrow 1$  & $\left(\tfrac{1}{5},\tfrac{3}{5},\tfrac{1}{5}\right)$ & $\big|\left(-\tfrac{1}{5},0,\tfrac{1}{5} \right); \varnothing\rangle$\\
& $\left(8,0 \right)$ & $t \rightarrow 1$  & $\left(\tfrac{3}{5},\tfrac{4}{5},\tfrac{3}{5}\right)$ & $\big|\left(-\tfrac{1}{10},0,\tfrac{1}{10} \right); \varnothing\rangle$\\
\hdashline
& $\left(10,0\right)$ & round		    & $\left(0,0,0\right)$ 					 &  $\big|\left(0,0,0\right); \varnothing\rangle$\\
\hdashline
& $\left(1,3 \right)$ & mixed		    & $\left(\tfrac{2}{5},0,\tfrac{3}{5}\right)$ 		 &$\big|\left(\tfrac{3}{10},-\tfrac{1}{10},-\tfrac{1}{5}\right); e^{2,+}_{0}e^{2,-}_{0} = -\tfrac{1}{2} \rangle$\\
& $\left(3,1 \right)$ & mixed		    & $\left(0,\tfrac{3}{5},\tfrac{2}{5}\right)$ 		&$\big|\left(-\tfrac{1}{10},-\tfrac{1}{5},\tfrac{3}{10}\right); e^{1,+}_{0}e^{1,-}_{0} = -\tfrac{1}{2}\rangle$\\
& $\left(1,1 \right)$ & mixed		    & $\left(\tfrac{3}{5},\tfrac{2}{5},0\right)$ 		&$\big|\left(-\tfrac{1}{5},\tfrac{3}{10},-\tfrac{1}{10}\right); e^{3,+}_{0}e^{3,-}_{0} = -\tfrac{1}{2} \rangle$\\
& $\left(2,4 \right)$ & mixed		    & $\left(0,\tfrac{2}{5},\tfrac{3}{5}\right)$ 		&$\big|\left( \tfrac{1}{10},-\tfrac{3}{10},-\tfrac{1}{10} \right); e^{1,+}_{0}e^{1,-}_{0} = -\tfrac{1}{2} \rangle$\\
& $\left(4,2 \right)$ & mixed		    & $\left(\tfrac{3}{5},0,\tfrac{2}{5}\right)$ 		&$\big|\left( \tfrac{1}{5},\tfrac{1}{10},-\tfrac{3}{10} \right); e^{2,+}_{0}e^{2,-}_{0} = -\tfrac{1}{2} \rangle$\\
& $\left(4,4 \right)$ & mixed		    & $\left(\tfrac{2}{5},\tfrac{3}{5},0\right)$ 		&$\big|\left( -\tfrac{3}{10},\tfrac{1}{5},\tfrac{1}{10} \right); e^{3,+}_{0}e^{3,-}_{0} = -\tfrac{1}{2} \rangle$\\
\hdashline
& $\left(2,2 \right)$ & mixed		    & $\left(\tfrac{1}{5},\tfrac{4}{5},0\right)$ 		&$\big|\left(-\tfrac{2}{5},\tfrac{1}{10},\tfrac{3}{10}\right); e^{3,+}_{0}e^{3,-}_{0} = -\tfrac{1}{2} \rangle$\\
& $\left(3,3 \right)$ & mixed		    & $\left(\tfrac{4}{5},\tfrac{1}{5},0\right)$ 		&$\big|\left(-\tfrac{1}{10},\tfrac{2}{5},-\tfrac{3}{10}\right); e^{3,+}_{0}e^{3,-}_{0} = -\tfrac{1}{2} \rangle$\\
\hline
L(6,1) 	
& $\left(2,0 \right),\left(4,2 \right)$ & $t \rightarrow 1$  & $\left(\tfrac{2}{6},\tfrac{2}{6},\tfrac{2}{6}\right)$ & $\big|\left(0,0,0 \right); \varnothing\rangle$\\
& $\left(4,0 \right),\left(2,4 \right)$ & $t \rightarrow 1$  & $\left(\tfrac{4}{6},\tfrac{4}{6},\tfrac{4}{6}\right)$ & $\big|\left(0,0,0 \right); \varnothing\rangle$\\
\hdashline
& $\left(3,1 \right),\left(5,3 \right),\left(1,5 \right)$ & $t \rightarrow 1$  & $\left(\tfrac{5}{6},\tfrac{5}{6},\tfrac{2}{6}\right)$ & $\big|\left(\tfrac{1}{4},-\tfrac{1}{4},0 \right); \varnothing\rangle$\\
\hdashline
& $\left(1,3 \right)$ & $t \rightarrow 1$  & $\left(\tfrac{1}{6},\tfrac{1}{6},\tfrac{4}{6}\right)$ & $\big|\left(\tfrac{1}{4},-\tfrac{1}{4},0 \right); \varnothing\rangle$\\
\end{tabular}}
\caption{We list here examples of the lens index of the $(A_1, D_4)$ Argyres-Douglas theory for a few value of $r$, and their precise identification with the twisted chiral algebra of $\widehat{\mathfrak{su}(3)}_{-\frac{3}{2}}$. The first two columns list the lens index data. We then specify the correct limit in the third column, which allows one to precisely match to a twisted character with $(\lambda_1,\lambda_2,\lambda_3)$ -- as given in the fourth column -- and the vacuum/highest-weight prescription in the final column.}\label{tablesu3}
\end{center}
\end{table}


\section{Discussion}\label{SecDiscussion}

In this paper we introduce a new ``observable" in the dictionary between (a protected sector of) four-dimensional $\cN=2$ superconformal field theories and two-dimensional chiral algebras. On the four-dimensional side, the observable is given by the lens index (in a particular limit) and on the two-dimensional side it corresponds to the character of a certain twisted module of the chiral algebra (or more properly the generalized vertex operator algebra).\footnote{Twisted modules can also be obtained in the presence of the surface defect in the four-dimensional superconformal field theory \cite{Cordova:2017mhb, BPRtwisted}.} We have demonstrated this by working out the precise dictionary for the cases of four-dimensional $\cN=2$ free theories as well as Argyres-Douglas theories with protected sector corresponding to Virasoro minimal models, $\widehat{\mathfrak{su}(2)}_{-\frac{4}{3}}$ and $\widehat{\mathfrak{su}(3)}_{-\frac{3}{2}}$ affine Kac-Moody algebras.

This is strongly suggestive that a similar correspondence between lens indices and twisted characters should hold for general four-dimensional $\cN=2$ superconformal field theories. However, we have observed in this paper that the precise identification is actually rather nontrivial; as discussed in the examples of $\widehat{\mathfrak{su}(2)}_{-\frac{4}{3}}$ and $\widehat{\mathfrak{su}(3)}_{-\frac{3}{2}}$ Kac-Moody algebras, the identification of the (well-defined) lens index with a two-dimensional twisted character requires the precise definition of the correct null states. It turns out that this requires us to impose a particular set of conditions on the vacuum/highest-weight state. For generic four-dimensional superconformal field theories, and their corresponding two-dimensional chiral algebras, the structure of the required vacuum conditions is not evident thus far. 

We further observed that there are in general three different types of limits, of which only two (the $t\to 1$ and $t\to q$ limit) give the precise two dimensional twisted characters. The third limit (essentially the Macdonald limit), which we called the ``mixed limit", only gives an identification with the refined (twisted) character, as introduced for the (untwisted) vacuum characters of Kac-Moody algebras in~\cite{Song:2016yfd}. We have further seen that the refined character correctly reproduces the Macdonald limit of the (lens) index away from any choice of limit. This further supports the claim that the notion of the refined character is well-defined more generally. It would be desirable to have an explicit derivation of the connection between the lens index and the twisted chiral algebra (or VOA). It would especially be interesting to have a top-down approach for identifying the choice of vacuum state. This problem may be related to finding a purely four-dimensional interpretation of the filtration \cite{feigin2009pbw,Li2005} used to define the refined character in the current paper.


\acknowledgments
We thank Shu-Heng Shao for collaboration at the early stage of this work. We would also like to thank Chris Beem, Ying-Hsuan Lin, Kazunobu Maruyoshi and Emily Nardoni for discussions. The work of MF is supported by the David and Ellen Lee Postdoctoral Scholarship and the U.S. Department of Energy, Office of Science, Office of High Energy Physics, under Award Number DE-SC0011632. The work of JS is supported in part by the Overseas Research Program for Young Scientists through Korea Institute for Advanced Study (KIAS).


\appendix

\section{Definition and review of the lens index} \label{app:LensIdx}

In this Appendix, we carefully define the $\cN=2$ superconformal lens index~\cite{Benini:2011nc} (see also~\cite{Alday:2013rs,Razamat:2013jxa,Razamat:2013opa}). At the end of the Appendix we also write down the ingredients to calculate the $\cN=1$ lens index, since we use them in the main text. 

\subsection{$\cN=2$ lens index}

The (round) superconformal index is defined as follows\footnote{We adopt the notation used in~\cite{Beem:2013sza}.}
\bea
\cI & = & \tr_{\cH} \left( -1 \right)^{F} \left( \frac{t}{ pq} \right)^{r} p^{j_{12}} q^{j_{34}} t^{R} \prod_i a_i^{f_i} \, e^{- \beta \delta_{2 \dot{-}}} \,,
\eea
where
\bea
 \delta_{2 \dot{-}} & = & \{\tilde Q_{2 \dot{-}}, \tilde Q^{\dagger}_{2 \dot{-}}\} \ = \ E - 2 j_2-2 R+r \,.
\eea
We have used the usual definitions, \ie~$j_1,j_2$ are the Cartan generators of the $SU(2)_1 \times SU(2)_2 \sim SO(4)$ isometry of $\mathbb{R}^{4}$, $j_{12} = j_2-j_1$ and $j_{34} = j_1+j_2$ correspond to rotations of the two planes $\mathbb{C}_{12} \oplus \mathbb{C}_{34}  = \mathbb{R}^{4}$, and $R,r$ are the Cartan generators of the superconformal R-symmetry group $SU(2)_R \times U(1)_r$. Finally the trace is taken over the Hilbert space of states on $S^{1} \times S^{3}$ for the round superconformal index. 

In the case of the lens index we modify the above definition to only allow for the trace over states in the Hilbert space of $S^{1} \times S^{3}/\mathbb{Z}_r$. Here the orbifold acts on the Hopf fiber via $\mathbb{Z}_r \subset U(1)_1 \subset SU(2)_1 \times SU(2)_2 \sim SO(4)$. The precise definition then reads
\bea
\cI_r & = & \tr_{\cH_r} \left( -1 \right)^{F} \left( \frac{t}{ pq} \right)^{r} p^{j_{12}} q^{j_{34}} t^{R} \prod_i a_i^{f_i} \, e^{- \beta \delta_{2 \dot{-}}} \,.
\eea

{\renewcommand{\arraystretch}{1.5}
\begin{table}[h!]
\begin{centering}
\begin{tabular}{|c|c|c|c|c|c|c|c|}
\hline
Letters & $  \quad E\quad $ & $\quad j_1 \quad$ & $\quad  j_2\quad$ & $\quad R \quad$ & $\quad r\quad$    & $\quad\quad \cI\quad\quad$ \\
  \hline \hline
$  \phi$ & 1 & 0 & 0 & 0 & -1   & $\frac{pq}{t}$ \\
  \hline
$  \lambda^{1} _{\pm}$ & $  \frac{3}{2}$ & $  \pm  \frac{1}{2}$ & $0$ & $  \frac{1}{2}$ & $-  \frac{1}{2}$   &  $-p$, $-q$ \tabularnewline
  \hline
$  \bar{\lambda}_{2 \dot{+}}$  & $  \frac{3}{2}$ & $0$ & $  \frac{1}{2}$ & $  \frac{1}{2}$ & $  \frac{1}{2}$  &  $-t$ \\
  \hline
$  \bar{F}_{\dot{+}\dot{+}}$ & $2$ & $0$ & $1$ & $0$ & $0$   &  $pq$ \\
  \hline
  $  \partial_{-\dot{+}}  \lambda^1_{+}+  \partial_{+\dot{+}}  \lambda^1_{-}=0$ & $  \frac{5}{2}$ & $0$ & $  \frac{1}{2}$ & $  \frac{1}{2}$ &
 $  -\frac{1}{2}$ & $pq$  \\
  \hline
\hline
$Q$ & $1$ & $0$ & $0$ & $  \frac{1}{2}$ & $0$  & $\sqrt{t}$ \\
  \hline
$  \bar{\psi}_{\dot{+}}$ & $  \frac{3}{2}$ & $0$ & $  \frac{1}{2}$ & $0$ & $-  \frac{1}{2}$   & $-\frac{pq}{\sqrt{t}}$ \\
  \hline
    \hline
$  \partial_{\pm\dot{+}}$ & $1$ & $  \pm  \frac{1}{2}$ & $  \frac{1}{2}$ & $0$ & $0$    & $p$, $q$ \\
\hline
\end{tabular}
\par  \end{centering}
  \caption{ \label{table:LetterIndex}
  Here the contributions to the index by the single letter operators are listed. The first section refers to the vector multiplet $\left( \phi , \bar\phi, \lambda^{\cI}_{\alpha}, \bar\lambda_{\cI \dot \alpha}, F_{\alpha\beta}, \bar{F}_{\dot \alpha \dot \beta}  \right)$ (including an equation of motion constraint) and the second to a half-hypermultiplet $\left( Q,\bar{Q}, \psi_{\alpha}, \bar{\psi}_{\alpha} \right)$. Finally in the last section refers to the spacetime derivatives.
}
\label{letters}
\end{table}}

Let us first focus on the free bifundamental hypermultiplet (\ie~in a bifundamental representation of $SU(N)\times SU(N)$). Without the lens action, this is simply given by
\bea
 \cI_{H} &= &  \mathrm{PE}\left[\frac{\sqrt{t} + \frac{p q}{\sqrt{t}}}{(1-p)(1-q)} (a+a^{-1}) \left( \sum_{i=1}^{N} x_i \right) \left( \sum_{i=1}^{N} y_i \right)\right] \nn\\
& = & \prod_{i,j=1}^{N} \Gamma\left( \sqrt{t} (a x_i y_j)^\pm ; p, q \right) \, ,
\eea
where $x_i, y_i$ are $SU(N)$ flavor fugacity, and $a$ is a $U(1)$ flavor fugacity, and PE stands for the plethystic exponential defined as
\bea
\mathrm{PE}\left[ f \left( \{x_i\} \right) \right] & = & \exp\left( \sum_{n=1}^{\infty} \frac{f \left(\{x_i^{n}\}\right)}{n}\right) \,.
\eea
We further used a short-hand notation where repeated signs implies the product of each combination, for instance $\G(a z^\pm; p, q) \equiv \G(a z; p, q) \G(a z^{-1}; p, q)$, and the elliptic Gamma function is defined as 
\begin{align}
\G(z; p, q) = \prod_{m, n \ge 0} \frac{1-z^{-1} p^{m+1} q^{n+1}}{1-z p^m q^n} \, .
\end{align}

Now let us turn to the lens index. The single letters contributing to the hypermultiplet index are of the form
\begin{align}
 \left( \partial_{\pm \dot{+}} \right)^{k} Q \,, \qquad  \left( \partial_{\pm \dot{+}} \right)^{\ell} \bar{\psi}_{\dot{+}} \,, \quad \text{with }\ k, \ell \in \mathbb{N} \,.
\end{align}
We denote by $j_1$ the momentum around the Hopf fiber of $S^{3}$ (we shall denote this curve $\gamma$ in the following). The orbifold now acts on the Hopf fiber by a rotation of $2\pi/r$. Hence to obtain the lens index we require states to survive this projection and thus we only include states with $j_1 = 0 \mod r$. Let us for now forget about the $SU(N)\times SU(N)$ symmetries and only focus on the $U(1)$ flavor symmetry. We can then introduce the fugacity as above and consider states (more generally) which transform by a nontrivial phase $e^{\frac{2 \pi \ii m}{r}} \in U(1)$, with $m \in \{0, \ldots, r-1\}$, by going around $\gamma$. In other words we are turning on non-trivial holonomy for this flavor symmetry group around the non-contractible cycle $\gamma$. Thus now the states/operators $\cO$ contributing to the index are supposed to satisfy the condition
\begin{align}
 j_1 \left( \cO \right) = m \mod r \,.
\end{align}
This argument can of course can be extended to any global (or gauge) symmetry. In particular if we have some symmetry $G$ (global or gauge -- see below), we also have to specify holonomy conditions for the fugacities around the cycle $\gamma$. In particular we have \emph{continuous} fugacities around the temporal $S^{1}$ cycle and \emph{discrete} fugacities around the nontrivial Hopf cycle $\gamma$. In other words, we have to specify commuting group elements $\alpha, \beta \in G$ such that
\be
 \alpha^{r} \ = \ 1 \,, \qquad [ \alpha, \beta] \ = \ 0 \,.
\ee
Up to conjugation $\alpha$ is then a discrete element which can be chosen to be in the maximal torus of $G$
\be
\alpha \ \sim \ \left\{ e^{2 \pi \ii m_i / r} \right\}_{i = 1}^{\mathrm{rk} \, G} \,,
\ee
where we have expanded $\alpha$ in terms of the generators of the maximal torus of $G$.

In the case of the bifundamental free hypermultiplet with $SU(N) \times SU(N) \times U(1)$, one can turn on different discrete fugacities for all three components. Hence in order to define the lens index we have to specify $m_i, \widetilde{m}_j, n \in \{0 , \ldots, N-1 \}$, with $\sum_{i} m_i = \sum_i \widetilde{m}_i = 0 \mod r$, along with the usual \emph{continuous} fugacities which are also present in the round sphere case. The discrete data tells us what kinds of projection conditions the fields have to satisfy under the orbifold action, \ie~by going around $\gamma$ the fields will have to come back to their original value up to the holonomy specified by $m_i, \widetilde{m}_i$ and $n$. Thus for each component of the fields, by expanding in terms of Cartan generators, we require that the $i^{\mathrm{th}}$ component satisfies the condition $j_1 \ = \ m_i \mod r$ and $j_1 \ = \ \widetilde{m}_i \mod r$ for the two $SU(N)$ parts, and similarly for the $U(1)$ part.~\footnote{If we want to gauge this symmetry we of course have to sum over all the possible such discrete holonomies, in the same way one has to sum over twisted sectors in a usual orbifold theory.} 

Now let us briefly discuss which operators precisely contribute to this index. To begin with, we shall set $(m_i+\widetilde{m}_j) = 0 \mod r$ for the corresponding components of the generator of the Cartan of $\mathfrak{su}(N)\times \mathfrak{su}(N)\times \mathfrak{u}(1)$. In that case we can more or less treat this as a $U(1)$, and we have no non-trivial holonomy around $\gamma$ (see above). In the round case, a particle $\hat{X}$ with given charges $j_{12}( \hat{X} ),j_{34}( \hat{X} ),R( \hat{X} )$ and $r( \hat{X} )$ contributes as follows to the index
\be
 X \sum_{i,j=0}^{\infty} p^{i}  q^{j} \ = \ X \sum_{k=0}^{\infty} \left( pq \right)^{k}\left\{ \sum_{\ell=0}^{\infty} \left( p^{\ell} + q^{\ell} \right) - 1\right\} \ = \ \frac{X}{1-pq} \left( \frac{1}{1-q}+ \frac{p}{1-p} \right) \,,
\ee
where 
\be
X \ = \ p^{j_{12}\left( \hat{X} \right)-r\left( \hat{X} \right)}q^{j_{34}\left( \hat{X} \right)-r\left( \hat{X} \right)} t^{r\left( \hat{X} \right)+R\left( \hat{X} \right)}\,.
\ee
Now the orbifold only acts on $j_1$, and thus in order to satisfy the condition $j_1 = 0 \mod r$ we require that only operators of the form
\be
\left( \partial_{+ \dot{+}} \right)^{k_1}\left( \partial_{- \dot{+}} \right)^{k_2} \hat{X} \,, \qquad \text{with} \qquad  k_1-k_2 + j_1 ( \hat{X}) \ = \ 0 \mod r
\ee
contribute to the index. The $ j_1 ( \hat{X})$ part is an overall shift and can be factored out into the $X$-part, and so we obtain for the single letter the following expression
\be
 X \sum_{\substack{i,j=0 \\  (i-j) \in r \mathbb{Z}}}^{\infty} p^{i}  q^{j} \ = \ X \sum_{k=0}^{\infty} \left( pq \right)^{k}\left\{ \sum_{\ell=0}^{\infty} \left( p^{\ell r} + q^{\ell r} \right) - 1\right\} \ = \ \frac{X}{1-pq} \left( \frac{1}{1-q^{r}}+ \frac{p^{r}}{1-p^{r}} \right) \,.
\ee
Now let us add some nontrivial holonomy around $\gamma$ in the Cartan of $SU(N)\times SU(N)$. We shall fix a single Abelian sector and add now the prescription 
\be
j_1 \ = \ - \underbrace{(m_i+\widetilde{m_j}+n)}_{m} \mod r \,.
\ee
Exactly as above, this means that now that $\left( k_1-k_2 \right) \in r \mathbb{Z} + m$ and $\left( k_2-k_1 \right) \in r \mathbb{Z} - m$ and thus our sum is restricted to terms like that. Consequently we obtain
\be
X \sum_{k=0}^{\infty} \left( pq \right)^{k}\left\{ \sum_{\ell=0}^{\infty} p^{\ell r+[m]_r} + \sum_{\ell=1}^{\infty} q^{\ell r-[m]_r} \right\} \ = \ \frac{X}{1-pq} \left( \frac{q^{[m]_r}}{1-q^{r}}+ \frac{p^{r-[m]_r}}{1-p^{r}} \right) \,.
\ee
Here we have introduced the notation
\bea
[x]_r & = & \left\{ k \in \{0, \ldots, r-1\} \ \big| \ k \ = \  x \mod r  \right\}\,,
\eea
which we use throughout the paper. Let us emphasize that we have only looked at a single $U(1)$ Cartan sector of the full symmetry group $SU(N)\times SU(N)$ and we will have to include all discrete fugacities $m_i + \widetilde{m}_j + n$ as well as continuous fugacities such as $a$, $x_i$ and $y_j$.

As an example, let us write down the lens index for a $\cN=2$ free hypermultiplet in a bifundamental representation of $SU(N) \times SU(N)$
\bea
 \cI_{H}
 & = &  \cI^{(0)}_{H} \, \mathrm{PE}\Bigg[ \sum_{\substack{i,j=1\\ s=\pm 1}}^{N}\frac{1}{1-pq}\left( \frac{p^{[s (m_i + \widetilde{m}_j+n)]_r}}{1-p^{r}} + \frac{q^{r-[s (m_i + \widetilde{m}_j+n)]_r}}{1-q^{r}} \right) \left( \sqrt{t} - \frac{pq}{\sqrt{t}} \right) (a x_i y_j)^{s} \bigg] \nn\\
 & =& \cI^{(0)}_{H} \,  \prod_{\substack{i,j=1\\ s= \pm 1}}^{N} 
 \Gamma\left( \sqrt{t} p^{[s (m_i + \widetilde{m}_j+n)]_r} \left( a x_i y_j  \right)^{s}; pq, p^{r} \right)\nn\\
&&\qquad\qquad\quad \times  \Gamma\left( \sqrt{t}  q^{r-[s (m_i + \widetilde{m}_j+n)]_r} \left( a x_i y_j  \right)^{s}; pq, q^{r} \right)\,.
\eea
Finally $\cI^{(0)}_{H}$ corresponds to an overall factor arising from the Casimir energy of the theory on $S^{1}\times S^{3}/\mathbb{Z}_r$. It is given by
\be
 \cI^{(0)}_{H} \ = \ \left( \frac{pq}{t} \right)^{\frac{1}{4}\left\{\sum_{i,j,s} [s(m_i +\widetilde{m}_j)]_r-\frac{1}{r} [s(m_i +\widetilde{m}_j)]_r^{2}\right\}} \, . 
\ee
This factor plays an important role by giving certain selection rules for states in the intermediate channels in the Macdonald limit ($p \to 0$), see for instance~\cite{Alday:2013rs}. It is also the most crucial piece in the Coulomb branch limit ($p, q, t \to 0, \frac{pq}{t}=u$) of the lens index~\cite{Gukov:2016lki, Fredrickson:2017yka}. 

Now let us turn our attention to the free vector multiplet. The arguments work very similar as for the free hypermultiplet. The single letter for a free vector of the round superconformal index is given by
\bea
 i_{V, \, {S^{3}}} &=& \frac{1}{1-pq} \left( \frac{1}{1-q}+ \frac{p}{1-p} \right) \left[\frac{pq}{t}- p - q + pq+pq - t\right] 
 \eea
By the same rule as before we find for the lens index
\be
 i_{V, \, S^{3}/\mathbb{Z}_r} \ =\ \frac{1}{1-pq} \left( \frac{q^{[m]_r}}{1-q^{r}}+ \frac{p^{r-[m]_r}}{1-p^{r}} \right) \left[\frac{pq}{t} +pq - t- 1 \right]  + \delta_{[m]_r, 0}\,.
\ee
Here we have to be slightly more careful, since $\lambda^{1}_{\pm}$ has nonzero $j_1$ charges and come with an equation of motion constraint. For $[m]_r \ \neq \ 0$ the only contribution to the index can come from $\left( \partial_{- \dot{+}} \lambda^{1}_{+} \right)$ and $\left( \partial_{+ \dot{+}} \lambda^{1}_{-} \right)$, which have zero $j_1$ charge. However they are related to one another by an equation of motion, whose part we have to subtract, thus the full single letter contribution is
\bea
&& (-p) \sum_{\substack{i, j \geq 0 \\ i-j+1 \in r \mathbb{Z} + m}} p^{i}q^{j} +(-q) \sum_{\substack{i, j \geq 0 \\ i-j-1 \in r \mathbb{Z} + m}} p^{i}q^{j} +pq \sum_{\substack{i, j \geq 0 \\ i-j \in r \mathbb{Z} + m}} p^{i}q^{j}  \nn\\
&& \ = \  \frac{-1}{1-pq} \left( \frac{q^{[m]_r}}{1-q^{r}}+ \frac{p^{r-[m]_r}}{1-p^{r}} \right) \,.
\eea
The same initial formula holds for $[m]_r  = 0$ however in evaluating one has to be a bit more careful, \ie
\bea
&&
(-p) \sum_{\substack{i, j \geq 0 \\ i-j+1 \in r \mathbb{Z} }} p^{i}q^{j} +(-q) \sum_{\substack{i, j \geq 0 \\ i-j-1 \in r \mathbb{Z}}} p^{i}q^{j} +pq \sum_{\substack{i, j \geq 0 \\ i-j \in r \mathbb{Z}}} p^{i}q^{j}  \nn\\
&& =  1- \frac{1}{1-pq} \left( \frac{1}{1-q^{r}}+ \frac{p^{r}}{1-p^{r}} \right) \,.
\eea
Thus as an example, we explicitly find for the free vector multiplet of a gauge theory $G=SU(N)$
\bea
\cI_{V, \mathbf{m}} & = & \cI_{V}^{(0)}  \mathrm{PE} \Bigg[ \sum_{i \neq j}\left( \frac{1}{1-pq} \left( \frac{q^{[m_{ij}]_r}}{1- q^{r}} +\frac{p^{r-[m_{ij}]_r}}{1-p^{r}} \right) \left( \frac{pq}{t} +pq - t - 1\right) + \delta_{[m_{ij}]_r,0} \right)\frac{a_i}{a_j}\Bigg]\nn\\
& = & \cI_{V}^{(0)} \left( \frac{(p^{r};p^{r})_\infty \, (p^{r};p^{r})_\infty}{\Gamma\left( t; pq, p^{r} \right) \, \Gamma\left( t q^{r}; pq, q^{r} \right)} \right)^{N-1}  \prod_{\substack{i \neq j \\ m_i = m_j}} (1 - z_i / z_j)^{-1} \nn\\
&& \times \prod_{\substack{i \neq j \\ s = 0,1}} \Gamma\left( t^{s} p^{[m_i - m_j]_r} z_i/z_j ; pq, p^{r} \right)\Gamma\left( t^{s} q^{r-[m_i - m_j]_r} z_i/z_j ; pq, q^{r} \right) \,,
\eea
where $m_{ij} = m_i - m_j$, and the zero-point contribution $\cI^{(0)}_{V}$ comes again from the Casimir energy and is given by
\bea
\cI^{(0)}_{V} & = & \left( \frac{pq}{t} \right)^{-\frac{1}{2} \left( \sum_{i,j=1}^{N} [m_{ij}]_r - \frac{1}{r} [m_{ij}]_r^{2} \right)} \,.
\eea

\paragraph{Gauging operation in the lens index.}

Let us mention the operation corresponding to ``gauging" a flavor symmetry. We start with some initial theory $\cT_{\mathrm{in}}$ with global symmetry group $G_f$, and we want to gauge a subset of the flavor symmetry group $G\subset G_{f}$. The lens index of the theory after gauging is schematically given by the following operation
\bea
\sum_{m_{\ell}=0}^{r-1} \frac{1}{\left| \cW_{m} \right|} \oint \left[\diff z\right]_{m} \, \cI_{V, m}^{G} \,  \cI_{\cT_{\mathrm{in}}} \left( (z,m),  \cdots  \right) \,,
\eea
where $\cI_{\cT_{\mathrm{in}}} \left( (a , m_a) \cdots  \right) $ denotes the index of the initial theory $\cT_{\mathrm{in}}$, which is ungauged, $\cI_{V, m}^{G}$ is a vector multiplet in the adjoint representation of $G$, and by $(a,m_a)$ we denote the totality of fugacities and discrete holonomies associated to the subgroup $G \subset G_{f}$. The sum is taken over all possible discrete holonomies $\left\{m_\ell\right\}_{\ell=1}^{\mathrm{rk}\left( G \right)}$.
Furthermore $\left| \cW_{m} \right|$ is the cardinality of the Weyl group of the subgroup $G_m$ of $G$ preserved after the introduction of the discrete holonomies $m$ (remember that they generically break the associated group). Finally by $\left[\diff z\right]_{m}$ we denote the Haar measure of the preserved subgroup $G_m \subset G$, more precisely
\bea
\left[\diff z \right]_{m} & = & \left[\prod_{i=1}^{\mathrm{rk } \, G} \frac{\diff z_i}{2\pi \ii}\right] \prod_{\alpha \in \Delta\cup \{0\}} \frac{1}{\left( 1- z^{\alpha} \right)^{\delta_{[\alpha(m)]_r}}} \,,
\eea
where $\Delta\cup \{0\}$ denotes all the roots (including the number $\mathrm{rk}\, ( G )$ of zero-roots) of $G$. This might look divergent, however the factors coming from the zero roots are cancelled by the vector multiplets that are gauged.

\paragraph{Macdonald limit.}

Let us briefly mention the appropriate Macdonald limit of the single letter indices. The Macdonald limit for the round index is defined by taking the $p \to 0$ limit. Similarly for the lens index the single letter indices simplify as follows
\bea
i_{H, \, S^{3}/\mathbb{Z}_r} & \rightarrow & 
\left(\delta_{m,0}+ \frac{q^{r-m}}{1-q^{r}} \right)\sqrt{t} \,, \\
 i_{V, \, S^{3}/\mathbb{Z}_r} 
 &\rightarrow& - \frac{q^{m}\left( t+1\right)}{1-q^{r}}  + \delta_{m, 0}\,. 
\eea
It is clear right away that these formulas precisely agree for the ones of the round sphere index if $m=0$ and $r=1$.

Generally, the naive Macdonald limit $p \to 0$ of the index vanishes due to the zero point energy contribution. However in order to end up with twisted modules for chiral algebras, we exclusively take the limit, after having removed the overall zero point energy piece.

\subsection{$\cN=1$ lens index}\label{N1indexdefs}

We use the formulas of the $\cN=1$ lens index in the main text and thus we briefly write down the appropriate ingredients here. The counting of states works the same way as reviewed above for the $\cN=2$ lens index and so we shall not review the derivation of the $\cN=1$ lens index here (for details see~\cite{Benini:2011nc,Razamat:2013opa}). For the chiral multiplet we obtain
\bea
 \cI_{\chi ,\, r} \left(m, u , R \right)
& = & \cI^{(0)}_{\chi, \, r} \left(m,u,R\right)\Gamma\left( \left( pq \right)^{\tfrac{R}{2}} q^{r-m} u ; q^{r}, pq \right)\Gamma\left( \left( pq \right)^{\tfrac{R}{2}} p^{m} u ; p^{r}, pq \right) \,,\nn\\
\eea
where
\bea
 \cI_{\chi, \, r}^{(0)} \left(m,u,R\right)& = & \left( (pq)^{\tfrac{1-R}{2}}u^{-1} \right)^{\tfrac{m(r-m)}{2r}} \left( \frac{p}{q} \right)^{\tfrac{m(r-m)(r-2m)}{12r}}
\eea
is the zero-point contribution, and $R$ is the R-charge of the chiral multiplet. As an example let us write down the lens index of a free $\cN=1$ hypermultiplet of R-charge $R$ in a bifundamental representation of $SU(N)\times SU(N)$
\bea
 \cI_{\chi ,\, r, \, \mathrm{bif}}^{} \big((u,m),(w,n),(a,o)\big) & = & 
\prod_{i,j=1}^{N}\prod_{s = \pm1}
 \cI_{\chi, \, r}^{(0)} \left( [s ( m_i + n_j +o) ]_r, (u_i w_j a )^{s} ,R \right)\nn\\
&&  \qquad \qquad \times 
 \Gamma\left( \left( pq \right)^{R/2} q^{r-[s ( m_i + n_j +o) ]_r} (u_i w_j a)^{s}  ; q^{r}, pq \right)\nn\\
&&  \qquad \qquad \times 
\Gamma\left( \left( pq \right)^{R/2} p^{[s ( m_i + n_j +o)]_r} (u_i w_j a)^{s}  ; p^{r}, pq \right) \,.\nn\\
\eea
Here $\vec{u},\vec{w}$ are the fugacities corresponding to the $SU(N)$ flavor symmetry and $\vec{m},\vec{n}$ are the corresponding (discrete) holonomies, \ie~$m_i,n_j \in \{0, \ldots, r-1\}$ for $1\leq i,j \leq N-1$, and $m_N=-\sum_{i=1}^{N-1}m_i$ as well as $n_N=-\sum_{i=1}^{N-1}n_i$. Similarly $a$ corresponds to the fugacity of the $U(1)$ flavor symmetry and $o\in\{0, \ldots , r-1\}$ its discrete holonomy.

Similarly for the lens index for a free $\cN=1$ vector multiplet we obtain
\bea
\cI_{\mathrm{V},r}^{} \left( u,m \right)
& = & \frac{\cI_{\mathrm{V}, \, r}^{(0)}\left( u,m \right)}{(1-u^{-1})^{\delta_{m,0}} \Gamma\left( q^{m} u^{-1} ; q^{r}, pq \right)\Gamma\left( p^{r-m} u^{-1} ; p^{r}, pq \right)} \,,
\eea
where
\bea
\cI_{\mathrm{V}, \, r}^{(0)} \left( u,m \right) & = & \left( (pq)^{1/2}u^{-1} \right)^{-\tfrac{m(r-m)}{2r}} \left( \frac{q}{p} \right)^{\tfrac{m(r-m)(r-2m)} {12r}}\,.
\eea
For instance for a free $\cN=1$ $SU(N)$ vector multiplet we have
\bea
\cI_{\mathrm{V},r}^{} \left(a,m \right) & = &
\prod_{i<j}
 \frac{\cI_{\mathrm{V}, \, r}^{(0)}\left(\tfrac{a_i}{a_j},[m_i-m_j]_r\right)}{(1-a_i/a_j)^{\delta_{m,0}} \Gamma\left( q^{[m_i-m_j]_r} \frac{a_j}{a_i} ; q^{r}, pq \right)\Gamma\left( p^{r-[m_i-m_j]_r} \frac{a_j}{a_i} ; p^{r}, pq \right)} \times\nn\\
&&\qquad  \times\frac{\cI_{\mathrm{V}, \, r}^{(0)} \left(a_j/a_i,[m_j-m_i]_r\right)}{ (1-a_j/a_i)^{\delta_{m,0}} \Gamma\left( q^{[m_j-m_i]_r} \frac{a_i}{a_j} ; q^{r}, pq \right)\Gamma\left( p^{r-[m_j-m_i]_r} \frac{a_i}{a_j} ; p^{r}, pq \right)} \,.\nn\\
\eea


\bibliography{refs}

\end{document}